\documentclass[preprint2]{aastex63}
\received{June 1, 2019}
\revised{January 10, 2019}
\accepted{\today}
\submitjournal{ApJ}

\shorttitle{The COM content in L1498}
\shortauthors{Jim\'enez-Serra et al.}
\graphicspath{{./}{figures/}}
\begin{document}

\title{The complex organic molecular content in the L1498 starless core}

\correspondingauthor{Izaskun Jim\'enez-Serra}
\email{ijimenez@cab.inta-csic.es}

\author[0000-0003-4493-8714]{Izaskun Jim\'enez-Serra}
\affiliation{Centro de Astrobiolog\'{\i}a (CSIC-INTA), Ctra. de Torrej\'on a Ajalvir km 4, 28850, Torrej\'on de Ardoz, Spain}

\author{Anton I. Vasyunin}
\affiliation{Ural Federal University, Kuybysheva st. 48, 620002, Ekaterinburg, Russian Federation}
%\affiliation{AAS Journals Associate Editor-in-Chief}

%\collaboration{1}{(AAS Journals Data Scientists collaboration)}

\author{Silvia Spezzano}
\author{Paola Caselli}
\affiliation{Max Planck Institute for Extraterrestrial Physics, Giessenbachstrasse 1, 85748, Garching, Germany}

\author{Giuliana Cosentino}
\affiliation{Chalmers University of Technology, SE41296, Gothenburg, Sweden}

\author{Serena Viti}
\affiliation{University of Leiden, Niels Bohrweg 2, 2333 CA, Leiden, The Netherlands}

%\author{Nuria Marcelino}
%\affiliation{Instituto de F\'{\i}sica Fundamental (IFF-€"CSIC), Serrano 123, 28006, Madrid, Spain}



\begin{abstract}

Observations carried out toward starless and pre-stellar cores have revealed that complex organic molecules are prevalent in these objects, but it is unclear what chemical processes are involved in their formation. Recently, it has been shown that complex organics are preferentially produced at an intermediate-density shell within the L1544 pre-stellar core at radial distances of $\sim$4000$\,$au with respect to the core center. However, the spatial distribution of complex organics has only been inferred toward this core and it remains unknown whether these species present a similar behaviour in other cores. We report high-sensitivity observations carried out toward two positions in the L1498 pre-stellar core, the dust peak and a position located at a distance of $\sim$11000$\,$au from the center of the core where the emission of CH$_3$OH peaks. Similarly to L1544, our observations reveal that small O-bearing molecules and N-bearing species are enhanced by factors $\sim$4-14 toward the outer shell of L1498. However, unlike L1544, large O-bearing organics such as CH$_3$CHO, CH$_3$OCH$_3$ or CH$_3$OCHO are not detected within our sensitivity limits. For N-bearing organics, these species are more abundant toward the outer shell of the L1498 pre-stellar core than toward the one in L1544. We propose that the differences observed between O-bearing and N-bearing species in L1498 and L1544 are due to the different physical structure of these cores, which in turn is a consequence of their evolutionary stage, with L1498 being younger than L1544.  

\end{abstract}



\section{Introduction}

Complex Organic Molecules (or COMs) are defined in Astrochemistry as carbon-bearing compounds with at least 6 atoms in their structure \citep[][]{herbst09}. COMs were initially detected toward hot sources with temperatures $\geq$100$\,$K such as massive hot cores \citep[][]{hollis00,hollis06,belloche08,belloche13} or low-mass warm cores \citep[i.e. hot corinos;][]{bottinelli04,jorgensen12}. For this reason, it was first postulated that COMs only form on dust grains via hydrogenation and atom addition reactions \citep[as for e.g. methanol; see][]{watanabe02,charnley95} or radical-radical reactions favoured by the heating from the central protostar when the temperatures get higher than 30$\,$K \citep[see][]{garrod08}. This hypothesis implied that COMs could only form via hydrogenation in the coldest environments of the ISM such as dense dark cloud cores (with T$\leq$10$\,$K), and that, if formed, they would not be seen in the gas phase. 

In the past decade, however, there has been a number of observational works that have changed our view of COM formation in the ISM. Large COM species such as propylene (CH$_2$CHCH$_3$), acetaldehyde (CH$_3$CHO), dimethyl ether (CH$_3$OCH$_3$) or methyl formate (CH$_3$OCHO) have been found in the gas-phase toward dark cloud cores and starless/pre-stellar cores \citep{marcelino07,oberg10,bacmann12,cernicharo12,vastel14,jimenez16,soma18,agundez19,yoshida19,scibelli20}. These observations have triggered a plethora of theoretical and experimental studies to understand COM formation at low temperatures \citep[see e.g.][]{rawlings13,vasyunin13,balucani15,ivlev15,ruaud15,chuang16,quenard18,shingledecker18,holdship19,jin20}, although no consensus has been reached yet. 

To gain insight into the processes involved in the formation of COMs in cold sources, \citet{jimenez16} measured the abundance profiles of complex organics toward the L1544 pre-stellar core by means of high-sensitivity single-dish observations. Two positions in the core were observed: the dust peak of L1544 with a visual extinction A$_{\rm v}$$\geq$60$\,$mag (the core's center); and the position where CH$_3$OH, the most abundant COM, peaks in L1544 \citep[{\it the methanol peak};][]{bizzocchi14,spezzano16}. The latter position is representative of an outer, lower-density shell located at a radius of 4000 au from the core's center and visual extinction A$_{\rm v}$$\sim$15-16$\,$mag. The high-sensitivity observations not only revealed the presence of large COMs such as CH$_3$OCHO, CH$_3$OCH$_3$, and CH$_2$CHCN in L1544 \citep[these species had remained elusive in previous campaigns; see][]{vastel14}, but also that they are factors $\sim$2-10 more abundant toward the position of the methanol peak than toward L1544's center \citep{jimenez16}. 

The modelling of the chemistry of O-bearing COMs in L1544 indeed predicts a peak in the radial abundance profiles of these species at a distance of $\sim$4000$\,$au, coinciding with the location of the methanol peak and with the enhancement of O-bearing COMs \citep{vasyunin17}. The enhancement of these COMs in the lower-density shell of L1544 is driven by \citep[][]{jimenez16}: i) the freezing-out of CO onto dust grains, which triggers an active grain surface chemistry; ii) the intermediate visual extinction at this shell (of $\sim$7-8$\,$mag)\footnote{Note that the visual extinction toward the methanol peak is A$_{\rm v}$$\sim$15-16$\,$mag but this is along the whole line-of-sight. The abundances in the models of \citet{vasyunin17} are calculated as a function of radius, and therefore the visual extinction in the model has to be half that measured along the line-of-sight \citep[i.e. A$_{\rm v}$$\sim$7-8$\,$mag; see][]{jimenez16}.}, which is sufficient to prevent the UV photo-dissociation of COMs by the external interstellar radiation field; and iii) the moderate H$_2$ gas number densities within the shell, which prevent the severe depletion of O-bearing COMs onto grains. All this suggests that O-bearing COMs are preferentially formed toward the low-density shells in starless/pre-stellar cores. For N-bearing COMs, the observations of \citet[][]{jimenez16} show that the abundances of these species are also enhanced toward the position of the methanol peak in L1544 by small factors (of 1.2-1.8), although the peak intensity of their lines show the opposite behaviour to those of O-bearing COMs: while the O-bearing COM emission is stronger toward the position of the methanol peak, the N-bearing COM lines are brighter toward the core's center. Unlike for O-bearing COMs, the formation mechanisms for N-bearing species in starless/pre-stellar cores remain vastly unexplored  \citep[see e.g.][]{fedoseev18}.

Until recently, the spatial distribution of O-bearing and N-bearing species in starless and pre-stellar cores had been studied only toward a few of these cores \citep[e.g. in L1544, L1498 and L1517B;][]{tafalla06,bizzocchi14,spezzano17}. However, \citet{scibelli20} have recently observed the emission from CH$_3$OH and CH$_3$CHO toward a large sample of starless and pre-stellar cores in the Taurus molecular cloud. Their results show that not only these species are prevalent across the sample (CH$_3$OH is detected in 100\% of the sources, while CH$_3$CHO is detected in 70\% of them), but also that the emission of CH$_3$OH is extended with morphologies similar to those found toward other cores such as L1544, L1498 or L1517B\footnote{\citet{scibelli20} mapped only seven cores from their sample in the Taurus molecular complex. Despite the poor angular resolution of their observations (beam of $\sim$1 arcmin), they find indications for chemical differentiation of CH$_3$OH toward all the mapped cores.}. However, it remains unclear whether all starless/pre-stellar cores present the same level of chemical complexity regardless of their evolutionary stage or environment \citep[see e.g.][]{lattanzi20}, or whether COMs and their precursors are distributed in all cores in a similar fashion to that observed in L1544.

The L1498 starless core belongs to the Taurus molecular complex and it is located at a distance of 130 pc \citep{galli19}. By analysing the kinematics of the high-density gas, \citet{tafalla04} found that the core is accreting material from the surrounding cloud as revealed by clear infall asymmetries in the CS line profiles. However, the N$_2$H$^+$ emission does not show any infall signatures, which implies that the core itself is not undergoing gravitational collapse. Compared to pre-stellar (gravitationally-collapsing) cores such as L1544, L1498 shows a much flatter radial density distribution \citep[][]{tafalla04}, and its deuterium fractionation [determined as N(N$_2$D$^+$)/N(N$_2$H$^+$)] is only 4\%, significantly lower than that found in L1544 \citep[$\sim$23-25\%;][]{crapsi05,redaelli19}. This lower deuterium fractionation is expected since it is the result of the lower CO depletion in L1498. Therefore, L1498 is likely at an earlier stage of evolution than L1544.

As L1544, L1498 also presents a methanol-rich shell. However, its methanol peak emission is located at distances further away from the center of the core \citep[$\sim$11000$\,$au in L1498 vs. $\sim$4000$\,$au in L1544;][]{tafalla06}\footnote{Note that the L1498 core morphology is elongated and the methanol shell presents radii between values $r_{\rm min}$=7800$\,$au and $r_{\rm max}$=13000$\,$au.}. The visual extinction toward this position is also lower (A$_{\rm v}$$\sim$7-8$\,$mag; see below) than toward the methanol peak in L1544 (A$_{\rm v}$$\sim$15-16$\,$mag), which suggests that the behaviour of O-bearing and N-bearing COMs, and of their precursors, may be different from that found in L1544.

In this paper, we present deep observations of O-bearing and N-bearing COMs toward two positions in the L1498 starless core. The goal is to not only establish the level of chemical complexity in this core, but to understand the formation mechanisms of COMs in cold cores as a function of evolution. The manuscript is organised as follows. Section$\,$\ref{obs} describes the observations, while Section$\,$\ref{results} reports the results on the analysis of the COM and COM precursor emission for their excitation, column densities and molecular abundances. Section$\,$\ref{comparison} compares the abundances of the COMs and precursors measured toward L1498 with those obtained toward L1544. In Section$\,$\ref{discussion}, we model the chemistry of the observed COMs and precursors, and compare the model predictions with the observations. Finally, in Section$\,$\ref{conclusions}, we summarise our conclusions. 

\section{Observations}
\label{obs}

%TABLE1------------------------------------------------------------------
\begin{deluxetable}{lcccc}
%\tablenum{1}
\tablecaption{Frequency ranges, velocity resolution ($\delta \rm v$) and rms noise level ($rms$) of our observations\label{tab:RMS}}
\tablewidth{0pt}
\tablehead{
\colhead{Frequency} & \colhead{$\delta \rm v$} & \multicolumn{2}{c}{$rms$} \\
\colhead{(GHz)} & \colhead{(km s$^{-1}$)} & \multicolumn{2}{c}{(mK)} \\ \cline{3-4}
\colhead{} & \colhead{} & Dust peak & CH$_3$OH peak
}
%\decimalcolnumbers
\startdata
78.1-80.0 & 0.19 & 4.9 & 3.8 \\
81.4-83.3 & 0.18 & 4.4 & 3.4 \\
83.4-85.3 & 0.18 & 2.2 & 2.2 \\
86.7-88.5 & 0.17 & 2.0 & 1.8 \\
93.8-95.7 & 0.16 & 4.7 & 3.7 \\
95.7-97.2 & 0.16 & 7.2 & 8.4 \\
97.2-99.0 & 0.15 & 4.9 & 4.0 \\
99.1-100.9 & 0.15 & 2.3 & 2.1 \\
102.4-104.2 & 0.15 & 2.9 & 2.6 \\
109.2-111.03 & 0.14 & 15.3 & 13.5 \\
\enddata
%\tablecomments{The rms noise level for the frequency ranges 95.7-97.2 and 109.2-111.03 are higher than for the rest of ranges, because they were sufficient to detect the CH$_3$OH and CH$_3$CN lines used in this study with high signal-to-noise ratios.}
\end{deluxetable}
%---------------------------

\begin{figure*}
	\includegraphics[angle=270,width=2.0\columnwidth]{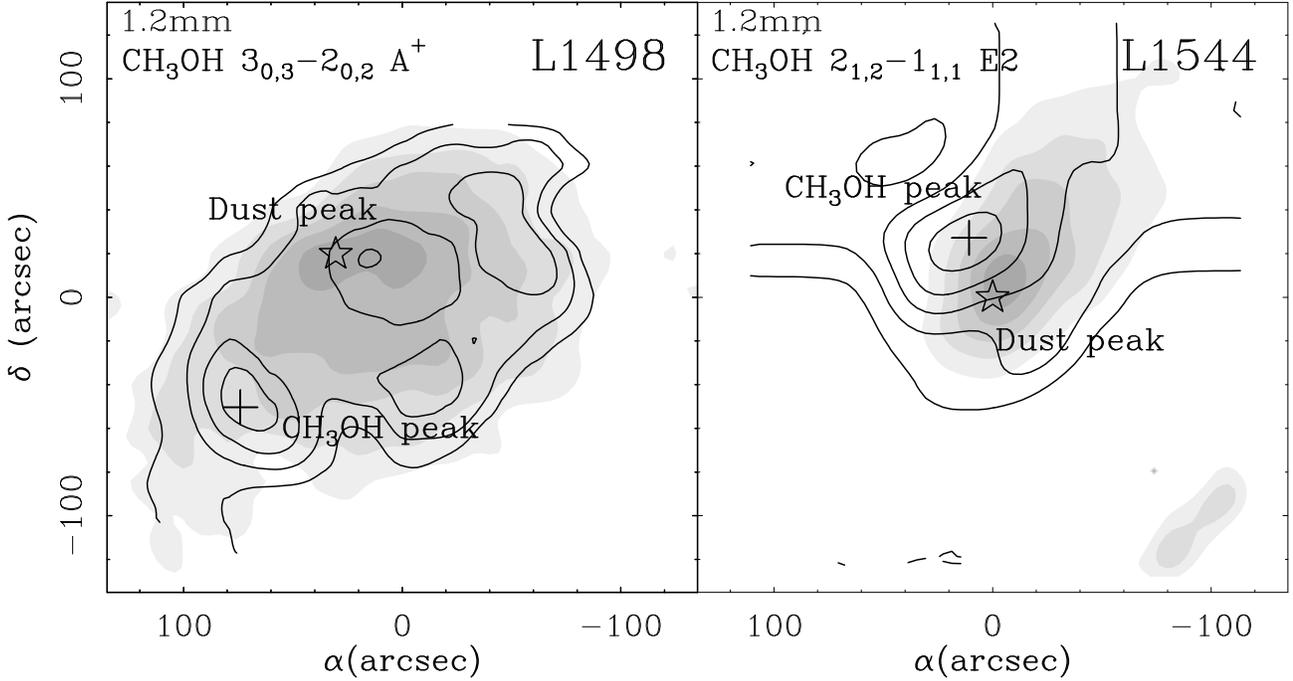}
    \caption{CH$_3$OH line integrated intensity maps (black contours) overlaid on the 1.2$\,$mm dust continuum maps (grey scale) obtained toward the L1498 (left panel) and L1544 (right panel) cold cores \citep[from][]{tafalla04,tafalla06,caselli02,spezzano17}. The grey scale indicates different percentage levels (15\%, 30\%, 45\%, 60\%, 75\% and 90\%) with respect to the 1.2$\,$mm dust continuum peak flux measured toward these cores (19.4 mJy/beam for L1498 and 224.8 mJy/beam for L1544). Contours correspond to the integrated intensity levels of 0.08, 0.12, 0.16 and 0.20 K$\,$km$\,$s$^{-1}$ for the CH$_3$OH 3$_{0,3}$$\rightarrow$2$_{0,2}$ A$^+$ in L1498, and to levels of 0.11, 0.19, 0.27, 0.35, and 0.43 K$\,$km$\,$s$^{-1}$ for the CH$_3$OH 2$_{1,2}$$\rightarrow$1$_{1,1}$ E2 in L1544. Stars indicate the location of the dust peaks in both cores while crosses show the coordinates of their methanol peaks. Note that the CH$_3$OH peak in L1544 is found at shorter radial distances ($\sim$4000$\,$au) than those associated with the CH$_3$OH peak in L1498 ($\sim$11000$\,$au).}
    \label{fig0}
\end{figure*}

The observations of the L1498 pre-stellar core were carried out at 3$\,$mm with the Instituto de Radioastronom\'{\i}a Milim\'etrica (IRAM) 30$\,$m telescope (Granada, Spain) between 15-17 March 2017 and 20-22 May 2017 (project number 094-16). Two positions were observed toward the L1498 core (see Figure$\,$\ref{fig0}): the dust continuum peak, with coordinates $\alpha$(J2000)=04$^h$10$^m$53.70$^s$, $\delta$(J2000)=25$^\circ$10$'$18.0$"$, and the position in the L1498 core where CH$_3$OH shows its peak emission, $\alpha$(J2000)=04$^h$10$^m$56.95$^s$, $\delta$(J2000)=25$^\circ$09$'$07.9$"$. The latter position (hereafter, the {\it methanol peak}) is indicated with a cross in Figure$\,$\ref{fig0} (left panel) and it is located $\sim$83$"$ away (or $\sim$11000$\,$au at a distance of 130 pc) from L1498's dust peak (shown as a star in the same panel). Similarly to what is seen toward the L1544 pre-stellar core, L1498's methanol peak appears towards a high-density tail associated with the core (Figure$\,$\ref{fig0}). However, the ring-like methanol structure detected toward L1498 is more homogeneous than that observed toward L1544 \citep[Figure$\,$\ref{fig0} and][]{bizzocchi14,tafalla06}. Note, however, that the methanol peak in L1544 (Figure$\,$\ref{fig0}, right panel) is found at shorter radial distances ($\sim$4000 au) that those associated with the methanol peak in L1498 ($\sim$11000$\,$au; left panel of Figure$\,$\ref{fig0}).

%FIG0 %%%%%%%%%%%%%%%%%%%%%%%%%%%%%
%%%%%%%%%%%%%%%%%%%%%%%%%%%%%%%%%%%%

The high-sensitivity 3$\,$mm spectra were obtained in frequency-switching mode using a frequency throw of 7.14$\,$MHz. The EMIR E090 receivers were tuned at 84.37$\,$GHz, 94.82$\,$GHz and 110.16$\,$GHz with rejections $\leq$10$\,$dB. The observed frequency ranges are shown in Table$\,$\ref{tab:RMS}. To avoid weak spurious features in the observed spectra, we carried out part of the observations by shifting slightly the central frequencies by $\pm$20$\,$MHz \citep[see also][]{jimenez16}. We used the narrow mode of the FTS spectrometer that provided a spectral resolution of 50$\,$kHz, equivalent to 0.15-0.18$\,$km$\,$s$^{-1}$ at 3$\,$mm. Typical system temperatures ranged between 75-110$\,$K and the telescope beam size was $\sim$24$"$-31$"$ between 79 and 101$\,$GHz. The spectra were calibrated in units of antenna temperature, T$^*_A$, and converted into main beam temperature, T$_{mb}$, by using a beam efficiency of 0.81 at 79-101$\,$GHz and of 0.78 at 102-111$\,$GHz. The rms noise level ranged between $\sim$2-15$\,$mK for the core's center position and between $\sim$1.8-14$\,$mK for the methanol peak (see Table$\,$\ref{tab:RMS}). The rms noise level for the frequencies between 95.7-97.2 and 109.2-111.03 were higher than for the rest of frequency ranges because they were sufficient to detect the observed CH$_3$OH and CH$_3$CN lines with high signal-to-noise ratios.

\section{Results}
\label{results}
\subsection{COMs and COM precursors spectra}
\label{spectra}

%FIG1 %%%%%%%%%%%%%%%%%%%%%%%%%%%%%
\begin{figure*}
	\includegraphics[angle=270,width=2.0\columnwidth]{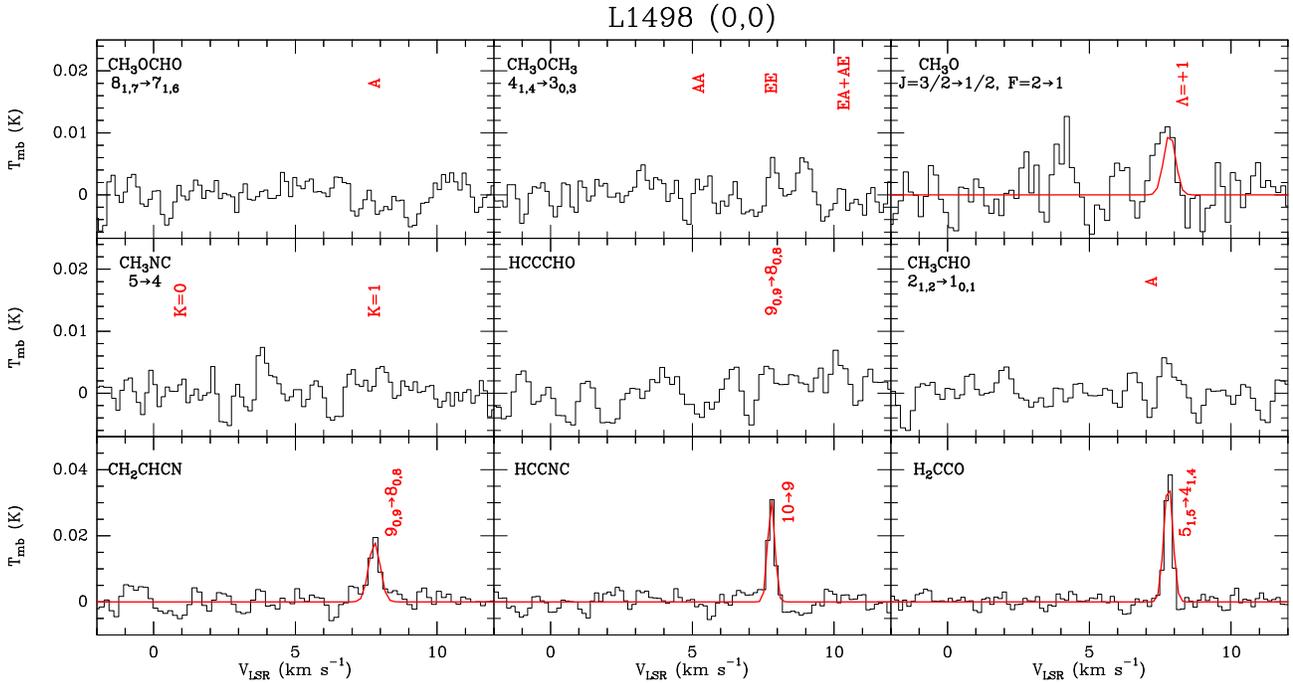}
    \caption{Sample of COM lines observed toward L1498's dust continuum peak. Red lines show the Gaussian line fits derived for the COMs and COM precursor transitions.}
    \label{fig1}
\end{figure*}
%%%%%%%%%%%%%%%%%%%%%%%%%%%%%%%%%%%%

Table$\,$\ref{tab:COMs-transitions} presents the COMs and COM precursor molecules observed toward the dust and methanol peaks of L1498, along with their measured transitions, derived line parameters and signal-to-noise ratio in integrated intensity (S/N). The identification of the lines has been carried out using the SLIM (Spectral Line Identification and Modelling) tool within the MADCUBA package\footnote{Madrid Data Cube Analysis on ImageJ is a software developed at the Center of Astrobiology (CAB) in Madrid; http://cab.intacsic.es/madcuba/Portada.html} \citep{martin19}. The spectroscopic entries are taken from the Cologne Database for Molecular Spectroscopy \citep[CDMS;][]{endres16} and the Jet Propulsion Laboratory \citep[JPL;][]{pickett98}.

\begin{figure*}
	\includegraphics[angle=270,width=2.0\columnwidth]{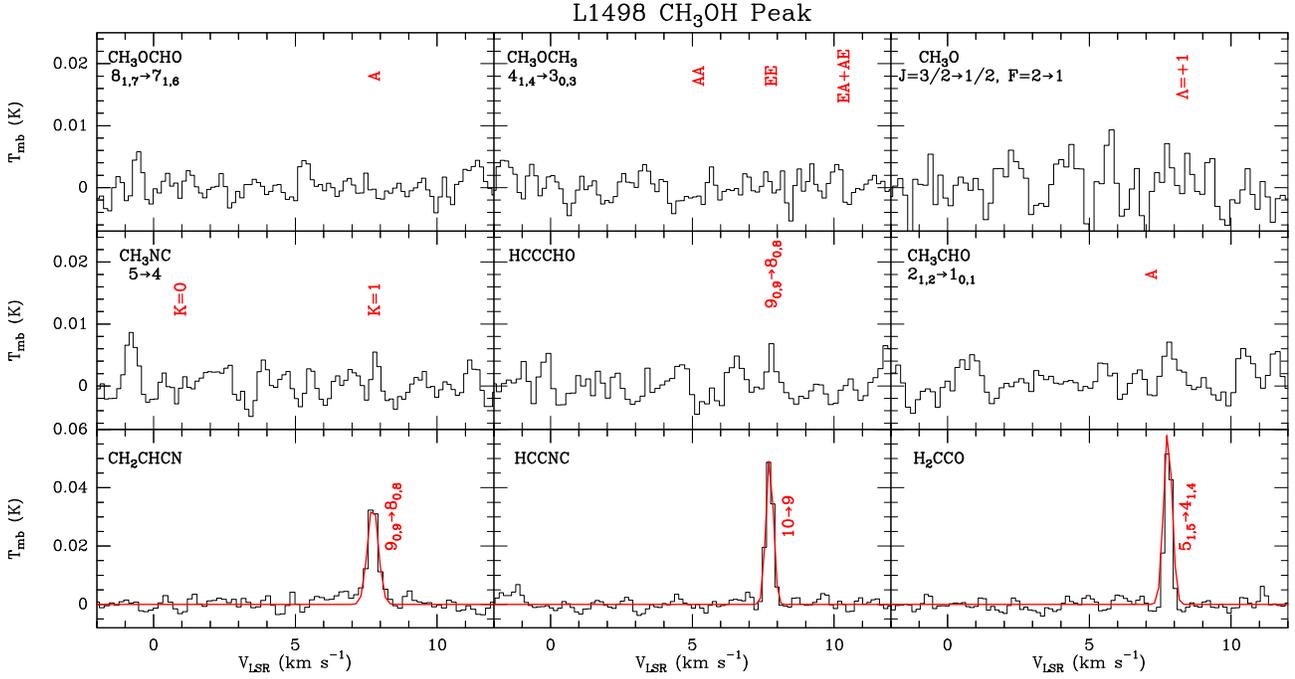}
    \caption{Sample of COM lines observed toward the position of the methanol peak in L1498. Red lines show the Gaussian line fits derived for the COMs and COM precursor transitions.}
    \label{fig2}
\end{figure*}

Our observations have covered both O-bearing and N-bearing COMs and COM precursors. For O-bearing species, we have targeted methanol (CH$_3$OH), methoxy (CH$_3$O), CCCO, ketene (H$_2$CCO), formic acid (t-HCOOH), acetaldehyde (CH$_3$CHO), methyl formate (CH$_3$OCHO), dimethylether (CH$_3$OCH$_3$), cyclopropenone (c-C$_3$H$_2$O), and propynal (HCCCHO). As N-bearing COMs and precursors, we have cyanoacetylene (HC$_3$N), isocyanoacetylene (HCCNC), vinyl cyanide (CH$_2$CHCN) and methyl isocyanide (CH$_3$NC). A sample of their observed transitions is presented in Figures$\,$\ref{fig1} and \ref{fig2} for the dust and methanol peaks of the L1498 starless core, respectively. 

Figures$\,$\ref{fig1} and \ref{fig2} show that species such as CH$_3$OH, CH$_3$O, CCCO, H$_2$CCO, HC$_3$N, HCCNC and CH$_2$CHCN, are detected either toward the dust or methanol peaks of the core, or toward both positions. The detected lines lie above the 3$\sigma$ level in integrated intensity, where 1$\sigma$ is calculated as $rms\times \sqrt{\Delta \rm v \times \delta \rm v}$, with $rms$ and $\delta$v being respectively the noise level and velocity resolution of the spectra (Table$\,$\ref{tab:RMS}), and $\Delta$v the linewidth of the line. We are confident about the detection of the lines because they all show central radial velocities $\sim$7.8$\,$km$\,$s$^{-1}$, the V$_{LSR}$ of the source. In addition, except for CCCO, we have covered at least two transitions for each detected species, which is enough for the correct identification of the molecular species since the level of line blending in starless/pre-stellar cores is expected to be low (linewidths of $\sim$0.3-0.4$\,$km$\,$s$^{-1}$; see Table$\,$\ref{tab:COMs-transitions}).

From Figures$\,$\ref{fig1} and \ref{fig2}, and from Table$\,$\ref{tab:COMs-transitions}, we find that the emission of the detected COMs and precursors tends to be brighter toward the position of the methanol peak with respect to the core's center except for CH$_3$O, which is only detected toward L1498's dust peak. Note, however, that the detected lines of CH$_3$O are weak (S/N=3.6; Table$\,$\ref{tab:COMs-transitions}). For the rest of molecules (t-HCOOH, CH$_3$CHO, CH$_3$OCHO, CH$_3$OCH$_3$, c-C$_3$H$_2$O, HCCCHO, and CH$_3$NC), we have not detected any feature. The upper limits have been obtained from those transitions expected to be the brightest at the T$_{ex}$ of the COM emission in the L1498 core (Section$\,$\ref{excitation}).  

%FIG2 %%%%%%%%%%%%%%%%%%%%%%%%%%%%%
%%%%%%%%%%%%%%%%%%%%%%%%%%%%%%%%%%%%

\subsection{Excitation analysis of COMs and COM precursors}
\label{excitation}

Except for CH$_3$OH (see below), the excitation temperature, T$_{ex}$, of the  COM species with at least two transitions covered in our observations (i.e. H$_2$CCO, CH$_2$CHCN, HC$_3$N and HCCNC; see Table$\,$\ref{tab:COMs-transitions}), has been determined using the SLIM package within MADCUBA. For CH$_3$O, the AUTOFIT tool did not converge because the targeted lines have similar upper level energies and therefore, the T$_{ex}$ was fixed to that obtained for H$_2$CCO, another COM precursor (T$_{ex}$=5.8$\pm$0.8$\,$K; Table$\,$\ref{tab:COMs-abun}). The same was done for CCCO, for which only one transition is available in our dataset. The T$_{ex}$ derived from the COM emission lie between 4.1 and 6.4$\,$K for the dust peak, and between 4.9 and 6.5$\,$K toward the methanol peak. These T$_{ex}$ are similar to those measured toward other starless/pre-stellar cores \citep[see][]{vastel14,bizzocchi14,jimenez16,punanova18,harju19,scibelli20}, and suggest sub-thermal excitation of the COM emission in the L1498 core given that the kinetic temperature of the gas is $\sim$10$\,$K within the inner $\sim$80" \citep[or 11000 au;][]{tafalla04}.  

The column densities of these COMs and COM precursors toward both the dust and methanol peaks in L1498 are reported in Table$\,$\ref{tab:COMs-abun}. The upper limits to the column densities of the molecular species not detected (e.g. CH$_3$OCHO or CH$_3$OCH$_3$) were calculated assuming T$_{ex}$=5.8$\,$K and are also reported in this Table.   

% LONG TABLE 2 -----------------------

%TABLE2------------------------------------------------------------------
\begin{longrotatetable}
\begin{deluxetable*}{lcccccccccc}
\tablecaption{COMs and COM precursors transitions covered in our L1498 observations and their derived line parameters\label{tab:COMs-transitions}}
\tablewidth{700pt}
\tabletypesize{\scriptsize}
\tablehead{
\colhead{Species} & \colhead{Line} & \colhead{Frequency} & \multicolumn{4}{c}{{\bf Dust peak}} &  \multicolumn{4}{c}{{\bf CH$_3$OH peak}} \\ \cline{4-11}
\colhead{} & \colhead{} & \colhead{} & \colhead{Area$^b$} & \colhead{$\Delta \rm v$} & \colhead{V$_{\rm LSR}$} & \colhead{S/N$^c$} & \colhead{Area$^b$} & \colhead{$\Delta \rm v$} & \colhead{V$_{\rm LSR}$} & \colhead{S/N$^c$} \\
\colhead{} & \colhead{} & \colhead{(MHz)} & \colhead{(mK km s$^{-1}$)} & \colhead{(km s$^{-1}$)} & \colhead{(km s$^{-1}$)} & \colhead{} & \colhead{(mK km s$^{-1}$)} & \colhead{(km s$^{-1}$)} & \colhead{(km s$^{-1}$)} & \colhead{}
} 
\startdata
CH$_3$OH & 2$_{0,2}$$\rightarrow$1$_{0,1}$ A & 96741.371 & 140.0(1.5)$^{e}$ & 0.228(0.007) &	7.796(0.004) & 93.3 & 160.0(1.8)$^{e}$ & 0.224(0.004) &	7.754(0.002) &  88.9 \\ 
	     & 2$_{1,2}$$\rightarrow$1$_{1,1}$ A & 95914.310 & $\leq$4.5 & $\ldots$ & $\ldots$ & $\ldots$ &  $\leq$5.4 & $\ldots$ & $\ldots$ & $\ldots$ \\
	     & 2$_{1,1}$$\rightarrow$1$_{1,0}$ A & 97582.798 & $\leq$4.5 & $\ldots$ & $\ldots$ & $\ldots$ &  $\leq$5.4 & $\ldots$ & $\ldots$ & $\ldots$ \\
	     & 2$_{1,2}$$\rightarrow$1$_{1,1}$ E & 96739.358 & 101.6(1.5) & 0.22(0.01) & 7.805(0.005) & 23.4 & 116.3(1.8) & 0.231(0.014) & 7.753(0.006) & 64.6 \\
	     & 2$_{0,2}$$\rightarrow$1$_{0,1}$ E & 96744.545 & 18.9(1.5) & 0.22(0.01) & 7.805(0.005) & 12.6 & 22.9(1.8) & 	0.231(0.014) &	7.753(0.006) &  12.7 \\
	     & 2$_{1,1}$$\rightarrow$1$_{1,0}$ E & 96755.501 & $\leq$4.5 & $\ldots$ & $\ldots$ & $\ldots$ & $\leq$5.4 & $\ldots$ & $\ldots$ & $\ldots$ \\ \hline
CH$_3$O$^a$ & F=1$\rightarrow$0, $\Lambda$=-1 & 82455.98 & $\leq$3.6 & $\ldots$ & $\ldots$ & $\ldots$ & $\leq$2.4 & $\ldots$ & $\ldots$ &  $\ldots$ \\
	& F=2$\rightarrow$1, $\Lambda$=-1 & 82458.25 & 4.3(1.2)$^c$ &	0.42(0.13) &	7.84(0.06) &	 3.6 & $\leq$2.4 & $\ldots$ & $\ldots$ & $\ldots$ \\
	& F=2$\rightarrow$1, $\Lambda$=+1 & 82471.82 & 4.3(1.2) &	0.42(0.13) & 7.84(0.06) & 3.6 & $\leq$2.4 & $\ldots$ & $\ldots$ & $\ldots$ \\
	& F=1$\rightarrow$0, $\Lambda$=+1 & 82524.18 & $\leq$3.6 & $\ldots$ & $\ldots$ & $\ldots$ & $\leq$2.4 & $\ldots$ & $\ldots$ &  $\ldots$\\ \hline
CCCO & 10$\rightarrow$9 & 96214.619 & 4.7(1.4) & 0.3(0.1) &	7.82(0.06) & 3.4 & 16.0(1.8) & 	0.43(0.08) & 7.82(0.03) & 8.9  \\ \hline
t-HCOOH & 1$_{1,1}$$\rightarrow$0$_{0,0}$ & 87926.877 & $\leq$1.5 & $\ldots$ & $\ldots$ & $\ldots$ & $\leq$1.2 & $\ldots$ & $\ldots$ &  $\ldots$\\ \hline
H$_2$CCO & 4$_{1,3}$$\rightarrow$3$_{1,2}$ & 81586.230 & 19.2(1.2) & 0.37(0.03) &	7.79(0.01) & 16 & 29.3(0.8) & 0.35(0.03) & 7.769(0.011) & 36.6 \\
              & 5$_{1,5}$$\rightarrow$4$_{1,4}$ & 100094.514 & 14.3(0.6) & 0.37(0.03) &7.79(0.01) & 23.8 & 22.3(0.8) 	& 0.35(0.03) & 7.769(0.011) &  27.9 \\ \hline	
CH$_3$OCHO & 7$_{2, 6}$$\rightarrow$6$_{2,5}$ A & 84454.754 & $\leq$1.8 & $\ldots$ & $\ldots$ & $\ldots$ & $\leq$1.5 & $\ldots$   & $\ldots$   & $\ldots$ \\ \hline
CH$_3$OCH$_3$ & 3$_{1,3}$$\rightarrow$2$_{0,2}$ EE & 82650.325 & $\leq$3.6 & $\ldots$ & $\ldots$ & $\ldots$ & $\leq$2.4 & $\ldots$   & $\ldots$   & $\ldots$ \\ \hline
CH$_3$CHO     & 5$_{0,5}$$\rightarrow$4$_{0,4}$ A  & 95963.465  & $\leq$5.3 & $\ldots$ & $\ldots$ & $\ldots$ & $\leq$5.4 & $\ldots$   & $\ldots$   & $\ldots$ \\ \hline
c-C$_3$H$_2$O & 6$_{1,6}$$\rightarrow$5$_{1,5}$    & 79483.520  & $\leq$4.2 & $\ldots$ & $\ldots$ & $\ldots$ & $\leq$2.7 & $\ldots$   & $\ldots$   & $\ldots$ \\ \hline
HCCCHO 	      & 9$_{0,9}$$\rightarrow$8$_{0,8}$    & 83775.842  & $\leq$1.8 & $\ldots$ & $\ldots$ & $\ldots$ & $\leq$1.5 & $\ldots$   & $\ldots$   & $\ldots$ \\ \hline
CH3CN & 6$_0$$\rightarrow$5$_0$  & 110383.500 & $\leq$9.7 & $\ldots$ & $\ldots$ & $\ldots$ & 25.7(3.1) & 0.32(0.05) & 7.72(0.02) & 8.3 \\
             &  6$_1$$\rightarrow$5$_1$  & 110381.372 & $\leq$9.7 & $\ldots$ & $\ldots$ & $\ldots$ & 17.2(3.1) & 0.32(0.05) &  7.72(0.02) & 5.5 \\
             &  6$_2$$\rightarrow$5$_2$  & 110374.989 & $\leq$9.7 & $\ldots$ & $\ldots$ & $\ldots$ & $\leq$9.3 &  $\ldots$ & $\ldots$ & $\ldots$ \\ \hline
CH$_3$NC      & 5$_0$$\rightarrow$4$_0$  & 100526.541 & $\leq$1.7 & $\ldots$ & $\ldots$ & $\ldots$ & $\leq$1.2 & $\ldots$   & $\ldots$   & $\ldots$ \\ \hline	
CH$_2$CHCN    & 9$_{0,9}$$\rightarrow$8$_{0,8}$    & 84946.000  & 6.3(0.6) & 0.36(0.14) & 7.78(0.06) & 10.5 & 13.0(1.5) & 	0.34(0.08) & 7.74(0.03) & 8.7 \\
			  & 9$_{1,9}$$\rightarrow$8$_{1,8}$    & 83207.505  & 4.4(1.2) & 0.36(0.14) & 7.78(0.06) & 3.7 & 8.9(1.5) & 	0.34(0.08) & 7.74(0.03) & 5.9 \\
	                   & 9$_{1,8}$$\rightarrow$8$_{1,7}$    & 87312.812  & 3.8(0.5) & 0.36(0.14) & 7.78(0.06) & 7.6 & 7.7(1.2) & 	0.34(0.08) & 7.74(0.03)    & 6.4 \\
	                   & 10$_{0,10}$$\rightarrow$9$_{0,9}$    & 94276.636  & $\leq$3.9 & $\ldots$ & $\ldots$ & $\ldots$ & 6.6(0.8) & 	0.34(0.08) & 7.74(0.03)    & 8.3 \\
	                   & 10$_{1,9}$$\rightarrow$9$_{1,8}$    & 96982.442  & $\leq$5.7 & $\ldots$ & $\ldots$ & $\ldots$ & $\leq$5.4 & $\ldots$ & $\ldots$   & $\ldots$ \\ \hline
HC$_3$N & 9$\rightarrow$8 &  81881.468 & 856.7(1.2) & 0.311(0.009) & 7.794(0.003) & 713.9 & 1102.2(0.8) & 	0.298(0.002) &	7.718(0.001) & 1377.8 \\
	     & 11$\rightarrow$10 &  100076.392 & 380.6(0.6) & 0.311(0.009) & 7.794(0.003) & 634.3 & 514.9(0.8) & 0.298(0.002) &	7.718(0.001) &  643.6 \\ \hline
HCCNC	      & 8$\rightarrow$7$^d$  & 79484.131  & 33.9(1.4) & 0.42(0.05) & 7.81(0.02) & 24.2 & 49.1(2.7) & 0.36(0.06) & 7.742(0.024) & 18.2 \\ 
	              & 10$\rightarrow$9$^d$  & 99354.250  & 11.0(0.6) & 0.42(0.05) & 7.81(0.02) & 18.3 & 17.7(0.4) & 0.36(0.06) & 7.742(0.024) & 44.3 \\ \hline
\enddata
\tablecomments{Line profiles were fitted using MADCUBA (see Section$\,$\ref{excitation}).$^{(a)}$ Hyperfine components of the N=1-0, K=0, J=3/2$\rightarrow$1/2 transition of CH$_3$O. $^{(b)}$ Upper limits are calculated as 3$\times$$rms$$\times\sqrt{\Delta \rm v \times \delta \rm v}$, with $rms$ the noise level, $\Delta$v the linewidth and $\delta$v the velocity resolution of the spectrum.
$^{(c)}$ This refers to the S/N ratio in integrated intensity area.
$^{(d)}$ Hyperfine structure not resolved. $^{(e)}$ Errors in the line area are calculated as $rms$$\times\sqrt{\Delta \rm v \times \delta \rm v}$.  
}
\end{deluxetable*}
\end{longrotatetable}
%------------------------------------------------------------------

For CH$_3$OH (A and E species), we used RADEX instead of MADCUBA \citep[][]{vandertak07} because MADCUBA performs the fitting of the molecular lines using the LTE approximation. We assume a kinetic temperature of T$_{kin}$=10$\,$K \citep[][]{tafalla04}. The fit of the CH$_3$OH lines was achieved considering an H$_2$ gas density of $\sim$4$\times$10$^4$$\,$cm$^{-3}$ (consistent with the one expected from the H$_2$ radial gas density profile derived by \citet{tafalla04} at a distance of $\sim$83$"$; see Section$\,$\ref{discussion}) and assuming column densities of 3.1$\times$10$^{12}$$\,$cm$^{-2}$/3.4$\times$10$^{12}$$\,$cm$^{-2}$ for the A/E species toward L1498's dust peak, and of 4.7$\times$10$^{12}$$\,$cm$^{-2}$/4.0$\times$10$^{12}$$\,$cm$^{-2}$ for the A/E species toward the methanol peak. The CH$_3$OH A/E abundance ratios are $\sim$0.9 and $\sim$1.2 toward the dust and methanol peaks, respectively. These values are similar to those derived toward L1544 and other molecular dark clouds such as L183, TMC-1 or B335 \citep[][]{friberg88,bizzocchi14}, and fall between the expected high temperature value of $\sim$1.0 and the one at 10$\,$K \citep[$\sim$0.7; see][]{wirstrom11}.

%TABLE3------------------------------------------------------------------
\begin{deluxetable*}{lccccccc}
%\tablenum{1}
\tablecaption{Excitation temperatures, column densities, and abundances of COMs and COM precursors  toward the dust and methanol peaks in L1498\label{tab:COMs-abun}}
\tablewidth{0pt}
\tablehead{
\colhead{Molecule} & \multicolumn{3}{c}{{\bf Dust Peak}} & \colhead{} & \multicolumn{3}{c}{{\bf Methanol Peak}} \\ \cline{2-4} \cline{6-8}
\colhead{} & \colhead{T$_{\rm ex}$} & \colhead{N$_{\rm obs}$ (cm$^{-2}$)} & \colhead{$\chi_{\rm obs}$$^{a}$} & \colhead{} & \colhead{T$_{\rm ex}$} & \colhead{N$_{\rm obs}$ (cm$^{-2}$)} & \colhead{$\chi_{\rm obs}$$^{a}$}
}
%\decimalcolnumbers
\startdata
        CH$_3$OH-A & 10$^{d}$ & 3.1$\times$10$^{12}$$^{d}$ & 8.9$\times$10$^{-11}$$^{d}$ & & 10$^{d}$ & 4.7$\times$10$^{12}$$^{d}$ & 6.5$\times$10$^{-10}$$^{d}$ \\
        CH$_3$OH-E & 10$^{d}$ & 3.4$\times$10$^{12}$$^{d}$ & 9.7$\times$10$^{-11}$$^{d}$ & & 10$^{d}$ & 4.0$\times$10$^{12}$$^{d}$ & 5.6$\times$10$^{-10}$$^{d}$ \\
        CH$_3$OH (Total) & $\ldots$ & 6.5$\times$10$^{12}$$^{d}$ & 1.9$\times$10$^{-10}$$^{d}$ & & $\ldots$ & 8.7$\times$10$^{12}$$^{d}$ & 1.2$\times$10$^{-9}$$^{d}$ \\
        CH$_3$O & 5.8$^{b}$ & (1.2$\pm$0.4)$\times$10$^{11}$ & (3.6$\pm$1.2)$\times$10$^{-12}$ & & 6.0$^{b}$ & $\leq$1.2$\times$10$^{11}$ & $\leq$1.7$\times$10$^{-11}$ \\
        CCCO & 5.8$^{b}$ & (3.9$\pm$1.5)$\times$10$^{11}$ & (1.2$\pm$0.5)$\times$10$^{-11}$ & & 6.0$^{b}$ & (1.2$\pm$0.2)$\times$10$^{12}$ & (1.7$\pm$0.3)$\times$10$^{-10}$ \\
        t-HCOOH & 5.8$^{b}$ & $\leq$4.6$\times$10$^{12}$ & $\leq$1.4$\times$10$^{-10}$ & & 6.0$^{b}$ & $\leq$6.6$\times$10$^{12}$ & $\leq$9.2$\times$10$^{-10}$ \\
        H$_2$CCO & 5.8$\pm$0.8$^{c}$ & (2.6$\pm$0.6)$\times$10$^{12}$ & (7.8$\pm$1.8)$\times$10$^{-11}$ & & 6.0$\pm$0.3$^{c}$ & (4.5$\pm$0.2)$\times$10$^{12}$ & (6.2$\pm$0.3)$\times$10$^{-10}$ \\
        CH$_3$OCHO & 5.8$^{b}$ & $\leq$1.3$\times$10$^{12}$ & $\leq$3.9$\times$10$^{-11}$ & & 6.0$^{b}$ & $\leq$1.1$\times$10$^{12}$ & $\leq$1.5$\times$10$^{-10}$ \\
        CH$_3$OCH$_3$ & 5.8$^{b}$ & $\leq$3.9$\times$10$^{11}$ & $\leq$1.2$\times$10$^{-11}$ & & 6.0$^{b}$ & $\leq$4.4$\times$10$^{11}$ & $\leq$6.1$\times$10$^{-11}$ \\
        CH$_3$CHO & 5.8$^{b}$ & $\leq$3.0$\times$10$^{11}$ & $\leq$9.0$\times$10$^{-12}$ & & 6.0$^{b}$ & $\leq$1.4$\times$10$^{11}$ & $\leq$1.9$\times$10$^{-11}$ \\
        c-C$_3$H$_2$O & 5.8$^{b}$ & $\leq$3.9$\times$10$^{10}$ & $\leq$1.2$\times$10$^{-12}$ & & 6.0$^{b}$ & $\leq$2.9$\times$10$^{10}$ & $\leq$4.0$\times$10$^{-12}$ \\
        HCCCHO & 5.8$^{b}$ & $\leq$2.0$\times$10$^{11}$ & $\leq$6.0$\times$10$^{-12}$ & & 6.0$^{b}$ & $\leq$2.0$\times$10$^{11}$ & $\leq$2.8$\times$10$^{-11}$ \\
        CH$_3$CN & 5.6$^{b}$ & $\leq$8.3$\times$10$^{10}$ & $\leq$2.4$\times$10$^{-12}$ & & 15.9$\pm$5.8 & (1.0$\pm$0.1)$\times$10$^{11}$ & (1.4$\pm$0.1)$\times$10$^{-11}$ \\
        CH$_3$NC & 5.6$^{b}$ & $\leq$1.0$\times$10$^{10}$ & $\leq$3.0$\times$10$^{-13}$ & & 6.0$^{b}$ & $\leq$7.8$\times$10$^{9}$ & $\leq$1.1$\times$10$^{-12}$ \\
        CH$_2$CHCN & 4.1$\pm$0.3$^{c}$ & (1.3$\pm$0.4)$\times$10$^{12}$ & (3.9$\pm$1.2)$\times$10$^{-11}$ & & 4.9$\pm$0.3$^{c}$ & (1.1$\pm$0.2)$\times$10$^{12}$ & (1.5$\pm$0.3)$\times$10$^{-10}$ \\
        HC$_3$N & 6.4$\pm$0.3$^{c}$ & (1.6$\pm$0.3)$\times$10$^{13}$ & (4.8$\pm$0.9)$\times$10$^{-10}$ & & 6.5$\pm$0.7$^{c}$ & (2.2$\pm$1.0)$\times$10$^{13}$ & (3.1$\pm$1.4)$\times$10$^{-9}$ \\
        HCCNC & 5.6$\pm$0.9$^{c}$ & (7.8$\pm$4.0)$\times$10$^{11}$ &(2.4$\pm$1.2)$\times$10$^{-11}$ & & 6.0$\pm$1.3$^{c}$ & (9.4$\pm$5.7)$\times$10$^{11}$ & (1.3$\pm$0.8)$\times$10$^{-10}$ \\ \hline
\enddata
\tablecomments{$^{(a)}$ Molecular abundances calculated using an H$_2$ column density of 3.5$\times$10$^{22}$ cm$^{-2}$ for the dust continuum peak and of 7.2$\times$10$^{21}$ cm$^{-2}$ for the position of the methanol peak (see Section$\,$\ref{abundance} for details). $^{(b)}$ Fitting solutions could only be obtained by fixing T$_{\rm ex}$ within the MADCUBA-SLIM tool \citep{martin19}. $^{(c)}$ Errors correspond to 1$\sigma$ uncertainties in MADCUBA. $^d$ The CH$_3$OH column densities have been calculated using RADEX \citep{vandertak07}. The value under T$_{ex}$ refers to the assumed kinetic temperature T$_{kin}$ in the RADEX calculations.}
\end{deluxetable*}
%------------------------------------------------------------------

%TABLE4------------------------------------------------------------------
\begin{deluxetable*}{lccccccc}
%\tablenum{1}
\tablecaption{Excitation temperatures, column densities, and abundances of CH$_3$OH, CCCO, H$_2$CCO, t-HCOOH and HC$_3$N toward the dust and methanol peaks in L1544\label{tab:COMs-abun-L1544}}
\tablewidth{0pt}
\tablehead{
\colhead{Molecule} & \multicolumn{3}{c}{{\bf Dust Peak}} & \colhead{} & \multicolumn{3}{c}{{\bf Methanol Peak}} \\ \cline{2-4} \cline{6-8}
\colhead{} & \colhead{T$_{\rm ex}$} & \colhead{N$_{\rm obs}$ (cm$^{-2}$)} & \colhead{$\chi_{\rm obs}$$^{a}$} & \colhead{} & \colhead{T$_{\rm ex}$} & \colhead{N$_{\rm obs}$ (cm$^{-2}$)} & \colhead{$\chi_{\rm obs}$$^{a}$}
}
%\decimalcolnumbers
\startdata
        CH$_3$OH-A & $\ldots$ & 0.7$\times$10$^{13}$$\,^{d}$ & 0.2$\times$10$^{-9}$$\,^{d}$ & & $\ldots$ & 2.4$\times$10$^{13}$$\,^{d}$ & 4$\times$10$^{-9}$$\,^{d}$ \\
        CH$_3$OH-E & $\ldots$ & 0.7$\times$10$^{13}$$\,^{d}$ & 0.2$\times$10$^{-9}$$\,^{d}$ & & $\ldots$ & 2.4$\times$10$^{13}$$\,^{d}$ & 4$\times$10$^{-9}$$\,^{d}$ \\
        CH$_3$OH (Total) & $\ldots$ & 1.4$\times$10$^{13}$$\,^{d}$ & 0.4$\times$10$^{-9}$$\,^{d}$ & & $\ldots$ & 4.8$\times$10$^{13}$$\,^{d}$ & 8$\times$10$^{-9}$$\,^{d}$ \\
        CCCO & 5.6$\pm$1.3$^{c}$ & (8.9$\pm$1.2)$\times$10$^{11}$$\,^{e}$ & (1.7$\pm$0.2)$\times$10$^{-11}$ & & 5.9$^{b}$ & (7.2$\pm$1.3)$\times$10$^{11}$$\,^{e}$ & (4.8$\pm$0.9)$\times$10$^{-11}$ \\
        t-HCOOH & 12.6$\pm$6.9$^{c}$ & (4.8$\pm$0.9)$\times$10$^{11}$$\,^{e}$ & (8.9$\pm$1.7)$\times$10$^{-12}$ & & 5.9$^{b}$ & (4.0$\pm$1.2)$\times$10$^{10}$$\,^{e}$ & (2.7$\pm$0.8)$\times$10$^{-12}$ \\
        H$_2$CCO & 4.9$\pm$0.2$\,^{f}$ & (1.5$\pm$0.2)$\times$10$^{13}$$\,^{f}$ & (2.8$\pm$0.4)$\times$10$^{-10}$ & & 5.9$\pm$0.8$^{c}$ & (9.8$\pm$2.3)$\times$10$^{12}$$\,^{e}$ & (6.5$\pm$1.5)$\times$10$^{-10}$ \\
        CH$_3$CN & 8.7$\pm$1.3$^{c}$ & (1.5$\pm$0.2)$\times$10$^{11}$$\,^{e}$ & (2.8$\pm$0.4)$\times$10$^{-12}$ & & 5.9$^{b}$ & $\leq$9.1$\times$10$^{10}$$\,^{e}$ & $\leq$6.1$\times$10$^{-12}$ \\
        HC$_3$N & 5.5$\pm$0.2$^{c}$ & (1.0$\pm$0.3)$\times$10$^{14}$$\,^{e}$ & (1.9$\pm$0.6)$\times$10$^{-9}$ & & 4.8$^{b}$ & (4.2$\pm$0.4)$\times$10$^{13}$$\,^{e}$ & (2.8$\pm$0.3)$\times$10$^{-9}$ \\ \hline
\enddata
\tablecomments{$^{(a)}$ Molecular abundances calculated using an H$_2$ column density of 5.4$\times$10$^{22}$ cm$^{-2}$ for the dust continuum peak and of 1.5$\times$10$^{22}$ cm$^{-2}$ for the position of the methanol peak \citep[][]{jimenez16}. $^{(b)}$ Fitting solutions could only be obtained by fixing T$_{\rm ex}$ within the MADCUBA-SLIM analysis tool \citep{martin19}. $^{(c)}$ Errors correspond to 1$\sigma$ uncertainties in MADCUBA. $^{(d)}$ Taken from the non-LTE analysis carried out by \citet{punanova18}. $^{(e)}$ Obtained using the data from \citet{jimenez16}. $^{(f)}$ Note that this molecule was also observed by \citet{vastel14} but they obtained a T$_{ex}$=27$\pm$1$\,$K. We  thus use the data from \citet{jimenez16}, which give a closer T$_{ex}$ to the ones inferred for the other species.}
\end{deluxetable*}
%------------------------------------------------------------------

\subsection{COM abundance profiles in L1498}
\label{abundance}

To calculate the abundances of the observed COMs and precursors, we need to determine the H$_2$ column density toward the dust and methanol peaks of L1498. For the core center, we use the H$_2$ column density of $\sim$3.5$\times$10$^{22}$$\,$cm$^{-2}$ inferred by \citet{tafalla04} from MAMBO 1.2$\,$mm data. The H$_2$ gas density profile of the L1498 starless core is flat within the inner 30$"$, which engulfs the radius associated with half the beam of the IRAM 30$\,$m observations at 3$\,$mm (radius of $\sim$13$"$). For the position of the methanol peak, we use the H$_2$ column density map of the Taurus molecular cloud obtained by \citet{spezzano16} from {\em Herschel} data at 250, 350 and 500$\,$$\mu m$ assuming a dust optical depth index of $\beta$=1.5. The H$_2$ column density toward L1498's methanol peak (i.e. at a distance of $\sim$83$"$ with respect to the core's center) is 7.2$\times$10$^{21}$$\,$cm$^{-2}$. While the H$_2$ column density of $\sim$3.5$\times$10$^{22}$$\,$cm$^{-2}$ toward the dust peak corresponds to a visual extinction of A$_{\rm v}$$\sim$37$\,$mag, the H$_2$ column density of 7.2$\times$10$^{21}$$\,$cm$^{-2}$ measured toward the methanol peak corresponds to a visual extinction of A$_{\rm v}$$\sim$7.7$\,$mag using the \citet[][]{bohlin78} formula (or N(H$_2$/A$_{\rm v}$)=0.94$\times$10$^{21}$$\,$cm$^{-2}$). 

Table$\,$\ref{tab:COMs-abun} reports the derived abundances of the COMs and COM precursors shown in Table$\,$\ref{tab:COMs-transitions}. Table$\,$\ref{tab:COMs-abun} reveals a general trend for these species to be enhanced toward the position of the methanol peak of L1498. The O-bearing species present the largest enhancements with CCCO and H$_2$CCO being, respectively, factors of $\sim$14 and $\sim$8 more abundant toward the methanol peak than toward the core center. For the N-bearing molecules, their abundances are enhanced by factors $\sim$4-6 toward the outer part of the L1498 starless core. This COM enhancement is similar to that reported toward the L1544 pre-stellar core \citep[][]{jimenez16}. In Section$\,$\ref{comparison}, we compare the COM and COM precursor abundance profiles measured toward these two cores, L1498 and L1544.

We note that the H$_2$ column density derived for the position of the dust peak but using {\em Herschel} data is $\sim$10$^{22}$$\,$cm$^{-2}$, a factor of $\sim$4 lower than that measured by MAMBO at 1.2$\,$mm. This is due to the fact that {\em Herschel} underestimates the total amount of dust present toward this region because it preferentially probes the dust grain population from the outer layers of the core \citep[][]{chacon-tanarro19} and the large beam can dilute possible centrally concentrated structures, as toward the center of L1544. This is the reason why we use the millimeter data for measuring the total H$_2$ column density toward the innermost regions of L1498. 
%Because of this difference once may think that the COMs enhancement found toward L1498's methanol peak is due to the choice of instrument providing the dust thermal emission flux (MAMBO vs. {\em Herschel}). However, 

For the position of the methanol peak, the H$_2$ column density measured using {\em Herschel} and MAMBO data are similar. Indeed, the H$_2$ column density obtained from the MAMBO 1.2$\,$mm map is $\sim$1.3$\times$10$^{22}$$\,$cm$^{-2}$, i.e. only a factor $\sim$1.8 higher than that derived using {\em Herschel} (7.2$\times$10$^{21}$$\,$cm$^{-2}$; see above). This factor is smaller than the COMs abundance enhancements measured in L1498's methanol peak with respect to the core center ($\sim$4-14), which implies that the observed increase in the abundance of COMs toward the outer regions of L1498 is real. We opt for the use of the {\em Herschel}-based H$_2$ column density toward L1498's methanol peak, because: i) the calculation and comparison with the COMs abundances toward L1544 is done in the same manner \citep[see][]{jimenez16}; ii) the {\em Herschel} beam of $\sim$35$"$ is closer to the one of the IRAM 30$\,$m observations ($\sim$30$"$ vs. $\sim$11$"$ of the MAMBO observations); and iii) the MAMBO observations toward the outer parts of the core may be affected by large-scale filtering produced by the chopping technique used in ground-based telescopes.

\section{Comparison with L1544}
\label{comparison}

\citet{vastel14} reported the abundances of the COMs and precursor species CH$_3$OH, CCCO, H$_2$CCO and t-HCOOH toward the dust peak of the L1544 pre-stellar core. In \citet{jimenez16}, two different positions in this core were observed but no abundances were provided for the latter COM precursor species. Therefore, in this Section, we take the abundances of CH$_3$OH (A and E species) from \citet[][who obtained a radial profile of the abundance of CH$_3$OH]{punanova18}, and calculate the abundances of CCCO, H$_2$CCO and t-HCOOH toward both the dust and methanol peaks of L1544 using the IRAM 30$\,$m data from \citet[][please refer to this article for the details on the observations]{jimenez16}. In this way, we can compare these results with those obtained toward the L1498 starless core. In our analysis, we also include the N-bearing molecules HC$_3$N and CH$_3$CN. 

Table$\,$\ref{tab:COMs-abun-L1544} collects the excitation temperatures (T$_{ex}$), column densities (N$_{obs}$), and molecular abundances of all these molecules measured toward the dust and methanol peaks of L1544. Note that the column densities derived for these species toward the dust peak of L1544 are very similar to those calculated independently by \citet{lattanzi20} for the same position within this core. 

\begin{figure*}
	\includegraphics[angle=0,width=2.0\columnwidth]{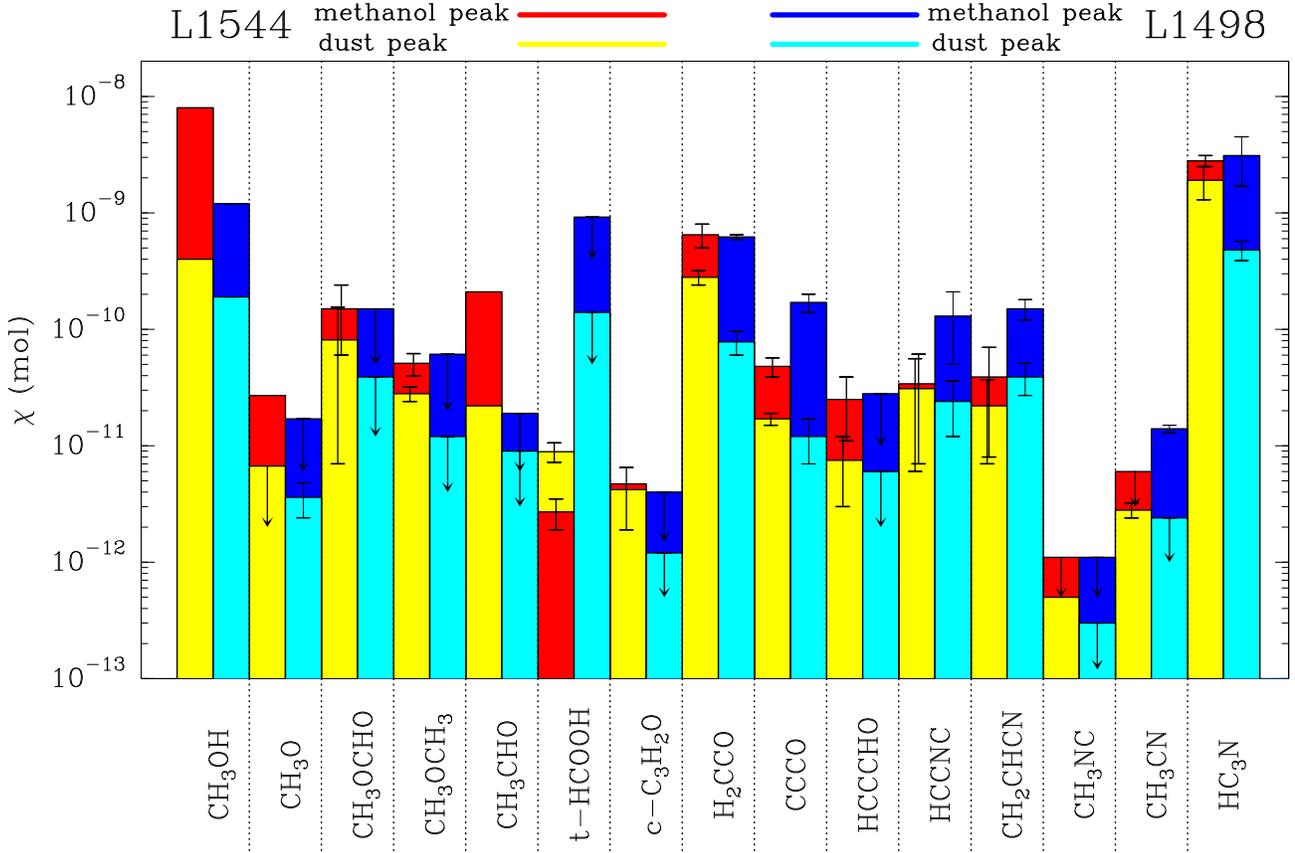}
    \caption{Histogram comparing the abundances of COMs and COM precursors measured toward the methanol and dust peaks in L1498 and L1544. Molecules are labeled at the bottom. Error bars correspond to 1$\sigma$ uncertainties in the column densities calculated by MADCUBA, and arrows indicate upper limits. Color legend is as follows: red - methanol peak in L1544; yellow - dust peak in L1544; dark blue - methanol peak in L1498; and cyan - dust peak in L1498. The abundances of the COMs and COM precursors have either been extracted from \citet{jimenez16} and \citet{punanova18}, or have been calculated from the data used in \citet[][]{jimenez16}.}
    \label{fig3}
\end{figure*}

In Figure$\,$\ref{fig3}, we present the comparison of the abundances of COMs and COM precursors measured toward the dust peaks and methanol peaks of the cold cores L1544 (yellow and red, respectively) and L1498 (cyan and dark blue, respectively). As mentioned in Section$\,$\ref{abundance}, all the detected COMs and COM precursors, except t-HCOOH in L1544, are enhanced toward the position of the methanol peaks in both L1498 and L1544. However, the abundances of O-bearing and N-bearing species are significantly different in L1498 with respect to L1544. While large O-bearing COMs such as CH$_3$CHO, CH$_3$OCHO or CH$_3$OCH$_3$ are clearly detected in L1544, we only report upper limits to the abundance of these molecules in L1498. Especially clear is the case of CH$_3$CHO, which is relatively abundant in L1544 (with an abundance of $\sim$2$\times$10$^{-10}$ toward the methanol peak; Figure$\,$\ref{fig3}), but it is not detected in L1498. Interestingly, the COM precursor CH$_3$O shows an opposite behaviour toward both cores: while CH$_3$O is clearly enhanced toward the methanol peak of L1544, this molecule is only detected toward the core center of L1498. However, note that the upper limit to the abundance of CH$_3$O toward L1498's methanol peak is still consistent with a possible enhancement of this species in the outer shells of the core. 

%FIG3 %%%%%%%%%%%%%%%%%%%%%%%%%%%%%
%%%%%%%%%%%%%%%%%%%%%%%%%%%%%%%%%%%%

For the N-bearing COMs and precursors, Figure$\,$\ref{fig3} shows that these species not only are in general more abundant toward L1498 than in L1544, but they also present larger enhancements toward L1498's methanol peak than toward the one in L1544 \citep[the enhancements are by factors 4-6 in L1498 vs. factors 1.2-1.8 in L1544; see Section$\,$\ref{abundance} and][]{jimenez16}. As discussed in Section$\,$\ref{discussion}, the observed differences in the abundance profiles of O-bearing and N-bearing species in L1498 and L1544 are due to the different physical structure of the cores, which in turn is likely associated with their evolutionary stage.

\section{Modelling the formation of COMs in L1498}
\label{discussion}

To understand the observed abundances of O-bearing and N-bearing COMs in L1498 and their differences with respect to L1544, we have modelled the chemistry of COMs using the 1D approach previously used in \citet[][]{vasyunin17} but applied to the case of L1498. The 1D density profile of the L1498 core was taken from \citet[][see their Table$\,$2]{tafalla04} and is:

\begin{equation}
    n(r) = \frac{0.94\times10^5}{1+(r(")/75")^{3.5}}\,\, \rm cm^{-3}.
\end{equation}

As temperature radial profile, we have used the T$_{kin}$ values derived by \citet[][]{tafalla04} from NH$_3$ observations for L1498 (see their Section$\,$3.1 and Figure$\,$4) and assumed a temperature radial distribution similar to the one employed by \citet{chacon-tanarro19} for the L1544 pre-stellar core. We also assume that the gas and dust temperatures are equal. The temperature profile is fitted as:

\begin{equation}
    T(r) = 13.0 - \frac{13.0 - 10.0}{1.0+(r('')/90'')^{1.7}}\,\, \rm K.
\end{equation}

\begin{figure}
	\includegraphics[angle=270,width=1.0\columnwidth]{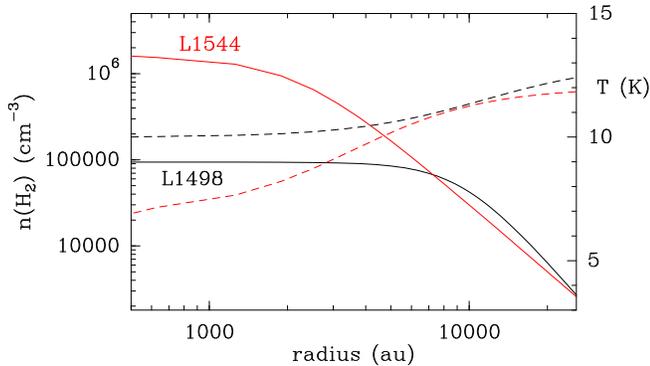}
    \caption{Radial distribution of the H$_2$ volume gas density (solid lines) and temperature (dashed lines) assumed for the L1498 starless core as a function of radius \citep[in black; see also Section$\,$\ref{discussion} and][]{tafalla04}. For comparison, we also show the physical structure of the L1544 pre-stellar core determined by \citet[][in red]{chacon-tanarro19}.}
    \label{fig4}
\end{figure}

The radial distribution of the H$_2$ volume gas density and temperature assumed for the L1498 starless core are shown in Figure$\,$\ref{fig4} \citep[see also][]{tafalla04}. For comparison, in this Figure we have also included the physical structure derived by \citet{chacon-tanarro19} for the L1544 pre-stellar core. From Figure$\,$\ref{fig4}, it is clear that the physical structure of both cores is different: L1498 shows a flatter gas density profile and lower peak H$_2$ gas density with respect to L1544, and its temperature at the center never gets below 10$\,$K as observed in L1544.

%FIG4 %%%%%%%%%%%%%%%%%%%%%%%%%%%%%
%%%%%%%%%%%%%%%%%%%%%%%%%%%%%%%%%%%%

Using the physical structure of L1498, the chemical evolution of the core is then calculated toward 150 radial points with 1$"$ spatial spacing. To simulate the chemical evolution at every radial point, we utilized the 0D gas-grain chemical model MONACO. This is a rate equations-based, three-phase (gas/ice surface/ice bulk) numerical model that provides the evolution of the fractional abundances of atomic and molecular species with time. The model is based on the work by~\citet[][]{vasyunin13} and was successfully applied to reproduce the abundances of O-bearing COMs in the prototypical L1544 pre-stellar core~\citep[][]{vasyunin17}. In this model, the precursors of both O-bearing and N-bearing COMs are formed on the surface of dust grains via hydrogenation. Once formed, a small fraction of these precursors is non-thermally desorbed from dust grains (see details below on the non-thermal processes assumed) and undergoes subsequent gas-phase reactions that produce the observed COMs \citep[see][]{vasyunin13,vasyunin17}.   

Before this study, minor updates had been introduced in the chemical network of the MONACO code as described in \citet[][see the reactions described in their Table$\,$2]{lattanzi20}. However, in our work, we have found an enhanced set of N-bearing species, which includes species such as CH$_{2}$CHCN, HCCNC and CH$_{3}$NC. Thus, we have further expanded our chemical network to account properly for the chemistry of those species. In particular, we have expanded the gas and grain chemistry of hydrocarbons (CxHy) following \citet[][]{hickson16}. This has enabled us to add the formation route of CH$_{2}$CHCN via the gas-phase reactions ${\rm N+C_{3}H_{5}\rightarrow CH_{2}CHCN+H_{2}}$ presented in the KIDA database, to the gas-phase and grain-surface routes already present in our network, namely the surface hydrogenation of HC$_{3}$N and the gas-phase reaction ${\rm CN + C_{2}H_{4} \rightarrow CH_2CHCN + H}$. We note, however, that the reaction between N and C$_{3}$H$_{5}$ has not been studied in detail, and it may not lead to CH$_{2}$CHCN as a product (J.-C. Loison; private communication). The chemical networks of HCCNC and CH$_{3}$NC were added following the approach described in \citet[][]{vastel19} and \citet[][]{willis20}.

It has been shown before that interstellar chemistry is an interplay between gas-phase and grain-surface chemical processes connected via the processes of accretion and desorption, and with surface chemistry being especially important for the formation of COMs \citep[see e.g.][]{herbst09,agundez13}. At the low temperatures of 10$\,$K typical of pre-stellar cores, desorption of species from dust grains to the gas phase requires additional energy which is either the result of exothermic surface reactions, or the result of the interaction of icy grain mantles with energetic particles such as cosmic rays or photons produced by cosmic rays. Although the details are still known poorly, many observational results such as widespread presence of methanol in cold gas imply non-zero efficiency of desorption from grains even at 10$\,$K. 

In our model, in addition to thermal evaporation, photodesorption, cosmic ray-induced desorption \citep[][]{hasegawa93} and reactive (chemical) desorption \citep[][]{minissale16} are included. While cosmic ray-induced desorption is known to have limited effects on the abundances of COMs, the case of reactive desorption is more complicated. Several theoretical and experimental studies of reactive desorption arrive to somewhat different conclusions on reactive desorption efficiency for various chemically reacting systems \citep[see][]{garrod07,minissale16,chuang18,fredon18}. The point of consensus is probably that the COMs with a large number of atoms ($>$6) do not desorb from grains via the mechanism of reactive desorption because the energy released during the exothermic surface reaction event likely gets redistributed and relaxed between the numerous internal degrees of freedom of a product molecule rather than breaking the bond between the product molecule and the surface. In this study, we use the parametrization of the efficiency of reactive desorption proposed in \citet[][]{minissale16}, the same as in \citet[][]{vasyunin17}. Note that a natural outcome of this parameterization is that the efficiency of chemical reactive desorption for large COMs (with more than 6 atoms such as e.g. CH$_2$CHCN) is negligible (see below).

Other assumptions made for the surface chemistry are also the same as in \citet[][]{vasyunin17}: the tunneling for surface diffusion of H and H$_{2}$ is enabled; and the diffusion-to-binding energy ratio E$_{\rm diff}$/E$_{\rm bind}$=0.3. The initial abundances of the species considered in the modeling were obtained from the final abundances of the simulation of a diffuse cloud model with A$_{V}$=2~mag, gas and dust temperatures of 20~K and gas density of 10$^{2}$~cm$^{-3}$ over 10$^{7}$~years using the $"$low metals'$"$ initial abundances EA1 from \citet[][]{wakelam08}.

Using the physical and chemical set-up described above, we performed the chemical modeling of the L1498 starless core over a time span of 10$^{6}$~years and compared the modeling results with our observations. We found that the best fit between the observed and the predicted fractional abundances of the O-bearing and N-bearing COMs (calculated using the modelled column densities convoluted with the 26'' IRAM 30m beam) is reached at an age of 9$\times$10$^{4}$~years for the L1498 starless core. The best-fit COMs abundances are presented in Table~\ref{tab:mod}, where we consider that the observed and predicted abundances are in agreement if both differ by less than a factor of 10. 

\begin{figure*}[ht!]
\includegraphics[angle=270,width=2.0\columnwidth]{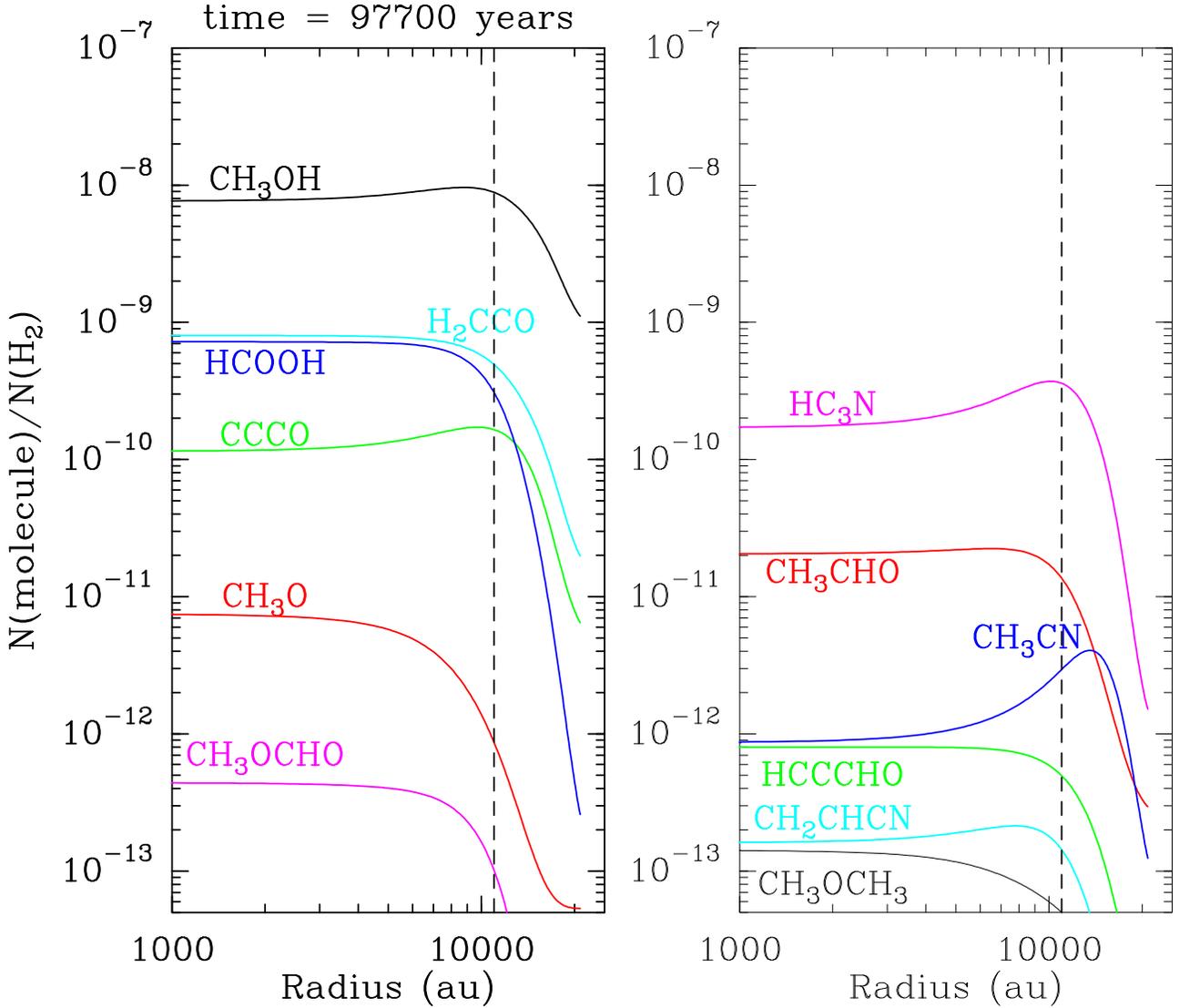}
\caption{Radial distributions of modeled fractional abundances of species considered in this study. Fractional abundances are derived as column density ratios convolved with a 26$"$ beam. A radius $r$=0 au corresponds to the location of the dust peak. Vertical dashed lines indicate the radius associated with the location of the methanol peak in L1498.\label{fig5}}
\end{figure*}

In Figure$\,$\ref{fig5}, we report the results of the abundance distribution of O-bearing and N-bearing COMs as a function of radius obtained for the L1498 starless core for our best-fit model. Interestingly, the predicted COMs abundances show a maximum in their distribution at a radial distance between 10000 au and 14000 au, which is consistent with the observed behaviour \citep[][]{tafalla06}. Note that this distance is three times further away than found toward L1544 \citep[$\sim$4000 au in L1544; see][]{vasyunin17} and it is a consequence of the flatter density distribution of the L1498 starless core. Our model also reproduces the lack of detections of O-bearing COMs such as CH$_3$OCHO and CH$_3$OCH$_3$ in L1498 as opposed to L1544. Indeed, the predicted abundances of these molecules for our L1498 model fall well below the value of 1$\times$10$^{-12}$, which is consistent with the measured upper limits for these species toward this core (see Table$\,$\ref{tab:COMs-abun} and Section$\,$\ref{results}). 

%FIG5 %%%%%%%%%%%%%%%%%%%%%%%%%%%%%
%%%%%%%%%%%%%%%%%%%%%%%%%%%%%%%%%%%%

Note, however, that the COM and COM precursor abundance enhancement predicted by the model at distances $\sim$11000 au, is less pronounced than that predicted by the same model for L1544. This is probably due to the fact that the H$_2$ gas density toward L1498's dust peak is moderate ($\sim$10$^5$$\,$cm$^{-3}$, i.e. two orders of magnitude lower than that measured for L1544) and thus, it does not induce such a strong molecular depletion in the L1498's model compared to L1544's one. The angular resolution of the MAMBO 1.2$\,$mm map of L1498 is however of  $\sim$11$"$ and, as in the case of L183 \citep[][]{lattanzi20}, we cannot rule out the possibility that within the innermost $\leq$2000$\,$au of L1498 the actual H$_2$ gas density is higher than $\sim$10$^5$$\,$cm$^{-3}$, as derived by \citet{tafalla04}. 

Despite this caveat, Table~\ref{tab:mod} shows that our model matches (overall) the abundances of most O-bearing and N-bearing COMs and COM precursors measured toward L1498, with the exception of CH$_3$OH and CH$_2$CHCN\footnote{H$_2$CCO shows a disagreement by a factor of just $\sim$10 toward the position of L1498's methanol peak (Table$\,$\ref{tab:mod}).}. For CH$_3$OH, our model overproduces this molecular species in the gas phase toward L1498's dust peak by a factor of $\sim$40 while it predicts an abundance within a factor of $\sim$7 for the position of the methanol peak. Note that this overproduction was already noted by \citet{punanova18} for the case of L1544, although in the case of L1498 it is more likely related to a milder depletion of the molecular gas toward the innermost regions of the core (see above).

%TABLE5---------------------------------------------------------------------------------------------------
\begin{deluxetable*}{lllccllc}
\tablecaption{Measured and modelled abundances of complex organic molecules in L1498\label{tab:mod}}
\tablewidth{0pt}
\tablehead{
\nocolhead{} & \multicolumn3c{Dust peak} & \nocolhead{} & \multicolumn3c{Methanol peak} \\
\cline{2-4}\cline{6-8}
\colhead{Species} & \colhead{${\rm N(X)/N(H_{2})_{obs}}$} & \colhead{${\rm N(X)/N(H_{2})_{mod}}$} & \colhead{Agreement} & \colhead{} & \colhead{${\rm N(X)/N(H_{2})_{obs}}$} & \colhead{${\rm N(X)/N(H_{2})_{mod}}$} & \colhead{Agreement} }
%\decimalcolnumbers
\startdata
CH$_{3}$OH        & 1.9$\times$10$^{-10}$ & 7.7$\times$10$^{-9}$ & $-$ & & 1.2$\times$10$^{-9}$ &        8.6$\times$10$^{-9}$ &      + \\
     CH$_{3}$O         &  3.6$\times$10$^{-12}$ & 7.5$\times$10$^{-12}$ & + & & $\leq$1.7$\times$10$^{-11}$ &        7.3$\times$10$^{-13}$ &      + \\
     C$_{3}$O          &    1.2$\times$10$^{-11}$ & 1.1$\times$10$^{-10}$ & + & & 1.7$\times$10$^{-10}$ &        1.6$\times$10$^{-10}$ &      +     \\
     HCOOH             &     $\leq$1.4$\times$10$^{-10}$ & 7.2$\times$10$^{-10}$ & + & & $\leq$9.2$\times$10$^{-10}$ &        2.7$\times$10$^{-10}$ &      + \\
     H$_{2}$CCO        &     7.8$\times$10$^{-11}$ & 8.0$\times$10$^{-10}$ & $-$ & & 6.2$\times$10$^{-10}$ &        4.6$\times$10$^{-10}$ &      +   \\
     HCOOCH$_{3}$      &      $\leq$3.9$\times$10$^{-11}$ & 4.4$\times$10$^{-13}$ & + & & $\leq$1.5$\times$10$^{-10}$ &        8.3$\times$10$^{-14}$ &      + \\
     CH$_{3}$OCH$_{3}$ &     $\leq$1.2$\times$10$^{-11}$ & 1.4$\times$10$^{-13}$ & + & & $\leq$6.1$\times$10$^{-11}$ &        4.7$\times$10$^{-14}$ &      + \\
     CH$_{3}$CHO       &      $\leq$9.0$\times$10$^{-12}$ & 2.1$\times$10$^{-11}$ & + & & $\leq$1.9$\times$10$^{-11}$ &        1.2$\times$10$^{-11}$ &      + \\
     HCCCHO            &      $\leq$7.2$\times$10$^{-12}$ & 8.0$\times$10$^{-13}$ & + & & $\leq$3.2$\times$10$^{-11}$ &        4.6$\times$10$^{-13}$ &      + \\
     CH$_3$CN         &      $\leq$2.4$\times$10$^{-12}$ & 8.7$\times$10$^{-13}$ & + & & 1.4$\times$10$^{-11}$ &  3.0$\times$10$^{-12}$   &  + \\
     CH$_{3}$NC        &      $\leq$3.0$\times$10$^{-13}$ & 8.7$\times$10$^{-13}$ & + & & $\leq$1.1$\times$10$^{-12}$ &        3.2$\times$10$^{-12}$ &      + \\
     CH$_{2}$CHCN      &      3.9$\times$10$^{-11}$ & 1.6$\times$10$^{-13}$ & $-$ & & 1.5e$\times$10$^{-10}$ &        1.3$\times$10$^{-13}$ & $-$  \\
     HC$_{3}$N         &      5.0$\times$10$^{-10}$ & 1.7$\times$10$^{-10}$ & + & & 3.2$\times$10$^{-9}$ &        3.5$\times$10$^{-10}$ &      +   \\
\enddata
%\tablecomments{UL stands for ``upper limit''}
\end{deluxetable*}
%----------------------------------------------------------------------------------------------------------------------

For CH$_{2}$CHCN, this molecule is firmly detected toward both positions in the L1498 starless core with abundances of (1.3-1.6)$\times$10$^{-10}$ (Table~\ref{tab:mod}), while the modelled values are up to three orders of magnitude lower. The low abundance of CH$_{2}$CHCN is maintained not only at the moment of the best agreement between the model and observations, but over the whole time span of the simulation: it never exceeds a few $\times$10$^{-12}$. This means that the proposed gas-phase formation routes of CH$_{2}$CHCN (i.e.  ${\rm N+C_{3}H_{5}\rightarrow CH_{2}CHCN+H_{2}}$ and ${\rm CN+C_{2}H_{4}\rightarrow CH_{2}CHCN+H}$, see above) are not sufficient to produce the observed amount of this species in the L1498 starless core. Alternatively, on grain surfaces CH$_2$CHCN forms mainly via the hydrogenation of HC$_3$N. However, this grain surface formation does not contribute significantly to the gas phase abundance of CH$_{2}$CHCN due to the near-zero efficiency of reactive desorption of this species from the grains \citep[of $\sim$10$^{-6}$, as calculated following the formalism of][]{minissale16}. This is in agreement with \citet[][]{garrod17} who also arrived to the conclusion that the modelled gas-phase abundance of CH$_{2}$CHCN does not exceed 10$^{-11}$ at low temperatures.

To reproduce the abundances of CH$_{2}$CHCN measured toward L1498, we have tested arbitrary values for the efficiency of reactive desorption of this species. An efficiency of 1\% is sufficient to reproduce the observed gas phase abundance of 1.5$\times$10$^{-10}$ for CH$_{2}$CHCN simultaneously with the rest of species. This implies that either chemical reactive desorption is somehow efficient for this molecule in particular, or that there is an additional mechanism responsible for the desorption of these large COMs besides chemical reactive desorption. The direct impact of cosmic rays on dust grains has been proposed to be a factor of 10 more efficient than chemical reactive desorption \citep[see][]{dartois19,dartois20}, which could increase the desorption efficiency of large COMs from grains. Note, however, that if this were the case, CH$_{2}$CHCN would have to survive on the grain surface over the period between its formation and independent desorption, which in our current model is unlikely because this molecule is further hydrogenated into ethyl cyanide (C$_{2}$H$_{5}$CN) on the grain surface. Further experimental and theoretical work on the grain chemistry and desorption mechanisms of N-bearing COMs such as C$_{2}$H$_{5}$CN at low temperatures is needed.

We finally note that the best-fit chemical age derived for L1498 is shorter than that for L1544 \citep[9$\times$10$^{4}$~years vs. 1.6$\times$10$^{5}$~years;][]{vasyunin17}. L1498 indeed shows a flatter density profile distribution than L1544 and its deuterium fractionation is consistent with the core being at an earlier evolutionary stage \citep{tafalla04,crapsi05}. The younger chemical age of L1498 is also supported by the smaller abundance difference found for methanol between its dust and methanol peaks, with respect to that measured toward L1544 (see Figure$\,$\ref{fig3}). This can be explained by the smaller degree of CO freeze-out in L1498, which in turn indicates a less evolved core for L1498 as compared to L1544. A shorter age would also explain why L1498 is accreting material from the molecular cloud onto the core, but does not show any infall signatures within the core itself yet \citep[][]{tafalla04}. 

\section{Conclusions}
\label{conclusions}

As shown in Section$\,$\ref{results}, our high-sensitivity observations carried out toward the L1498 starless core reveal that O-bearing COMs such as CH$_3$CHO, CH$_3$OCHO and CH$_3$OCH$_3$ are not detected neither toward the dust peak nor toward the outer regions of the core where CH$_3$OH peaks (i.e. at $\sim$10000 au from the core's center). This behaviour is in contrast with the one observed toward the L1544 pre-stellar core, where these large O-bearing COMs not only are detected toward this outer region, but their abundances present their maximum values at a distance of only $\sim$4000 au with respect to the center of the L1544 core.

For N-bearing COMs, these species are systematically enhanced toward the outer regions of the L1498 starless core in agreement with what was previously found toward L1544, and even in higher amounts. The distance at which the enhancement of the N-bearing COMs occurs in L1498 is, however, much larger than that found in L1544 ($\sim$10000 au vs $\sim$4000 au). 

We have investigated the origin of the differences observed in the behaviour of COMs between L1498 and L1544, by carrying out chemical modelling of both O-bearing and N-bearing COMs using the known physical structure of the L1498 starless core (Section$\,$\ref{discussion}). Our modelling results reveal that the enhancement of N-bearing COMs in the outer shells of L1498 at distances of $\sim$10000 au, is the expected outcome of the interplay between the gas-phase and the surface of dust grains given the flatter density radial profile of L1498 in comparison to the more evolved pre-stellar core L1544. Therefore, the different content in O-bearing and N-bearing complex organic material in starless and pre-stellar cores is a consequence not only of their physical structure but also, of their evolution.

\acknowledgments

We also thank the staff at the IRAM 30m telescope for their help and support during the observations. We would like to thank M. Tafalla for sharing his NH$_3$ T$_{kin}$ data results obtained toward L1498 and the dust and CH$_3$OH emission maps of this core shown in his paper \citet{tafalla06}. We also acknowledge the input provided by J.C. Loison on the viable formation reactions for vinyl cyanide and the constructive comments from an anonymous referee that helped to improve the original version of the manuscript. This work is based on observations carried out under projects number 094-16 and 013-20 with the IRAM 30m telescope. IRAM is supported by INSU/CNRS (France), MPG (Germany) and IGN (Spain). I.J.-S. has received partial support from the Spanish State Research Agency (AEI) project numbers PID2019-105552RB-C41 and MDM-2017-0737 Unidad de Excelencia $"$Mar\'{\i}a de Maeztu$"$- Centro de Astrobiolog\'{\i}a (CSIC-INTA), and from the Spanish National Research Council (CSIC) through the i-Link project number LINKA20353. The work by A.I.V. is supported by the Russian Ministry of Science and Higher Education via the State Assignment Contract FEUZ-2020-0038.


\vspace{5mm}
\facilities{IRAM 30m telescope}


\software{MADCUBA \citep{martin19},
                MONACO \citep{vasyunin17}      }



%\appendix

%\section{Appendix information}

\clearpage

% bibliography ---------------------




\begin{thebibliography}{}

\bibitem[Ag\'undez et al.(2019)]{agundez19}
Ag\'undez, M., Marcelino, N., Cernicharo, J., Roueff, E., \& Tafalla, M.  2019, A\&A, 625A, 147A

\bibitem[Ag\'undez \& Wakelam(2013)]{agundez13}
Ag\'undez, M., \& Wakelam, V. 2013, ChRv, 113, 8710A

\bibitem[Bacmann et al.(2012)]{bacmann12}
Bacmann, A., Taquet, V., Faure, A., Kahane, C., \& Ceccarelli, C. 2012, A\&A, 541, L12

\bibitem[Balucani et al.(2015)]{balucani15}
Balucani, N., Ceccarelli, C., \& Taquet, V. 2015, \mnras, 449, L16

\bibitem[\protect\citeauthoryear{Belloche et al.}{2013}]{belloche13}
Belloche, A., M\"uller, H., Menten, K., Schilke, P. \& Comito, C. 2013, A\&A, 559A, 47B

\bibitem[\protect\citeauthoryear{Belloche et al.}{2008}]{belloche08}
Belloche, A., et al. 2008, A\&A, 482, 179B

\bibitem[\protect\citeauthoryear{Bizzocchi et al.}{2014}]{bizzocchi14}
Bizzocchi, L., Caselli, P., Spezzano, S., \& Leonardo, E.  2014, A\&A, 569A, 27B

\bibitem[\protect\citeauthoryear{Bohlin et al.}{1978}]{bohlin78}
Bohlin, R. C., Savage, B. D., \& Drake, J. F. 1978, ApJ, 224, 132B

\bibitem[\protect\citeauthoryear{Bottinelli et al.}{2004}]{bottinelli04}
Bottinelli, S., et al. 2004, ApJ, 615, 354B

\bibitem[\protect\citeauthoryear{Caselli et al.}{2002}]{caselli02}
Caselli, P., Walmsley, C. M., Zucconi, A., Tafalla, M., Dore, L., \& Myers, P. C.  2002, ApJ, 565, 331C

\bibitem[\protect\citeauthoryear{Cernicharo et al.}{2012}]{cernicharo12}
Cernicharo, J., Marcelino, N., Roueff, E., Gerin, M., Jim\'enez-Escobar, A., Mu\~noz Caro, G. M. 2012, ApJ, 759L, 43C

\bibitem[\protect\citeauthoryear{Chac\'on-Tanarro et al.}{2019}]{chacon-tanarro19}
Chac\'on-Tanarro, A., et al. 2019, A\&A, 623A, 118C

\bibitem[\protect\citeauthoryear{Charnley et al.}{1995}]{charnley95}
Charnley, S. B., Kress, M. E., Tielens, A. G. G. M., \& Millar, T. J. 1995, ApJ, 448, 232C

\bibitem[Chuang et al.(2018)]{chuang18}	
Chuang, K. -J., Fedoseev, G., Qasim, D., Ioppolo, S., van Dishoeck, E. F., \& Linnartz, H. 2018, ApJ, 853, 102C

\bibitem[Chuang et al.(2016)]{chuang16}	
Chuang, K.-J., Fedoseev, G., Ioppolo, S., van Dishoeck, E. F., \& Linnartz, H. 2016, MNRAS, 455, 1702

\bibitem[\protect\citeauthoryear{Crapsi et al.}{2005}]{crapsi05}
Crapsi, A., et al., 2005, ApJ, 619, 379C
 
\bibitem[\protect\citeauthoryear{Dartois et al.}{2020}]{dartois20}
Dartois, E., Chabot, M., Bacmann, A., Boduch, P., Domaracka, A., \& Rothard, H. 2020, A\&A, 634A, 103D

\bibitem[\protect\citeauthoryear{Dartois et al.}{2019}]{dartois19}
Dartois, E., et al.  2019, A\&A, 627A, 55D

%\bibitem[\protect\citeauthoryear{Elias}{1978}]{elias78}
%Elias, J. H. 1978, ApJ, 224, 857

\bibitem[\protect\citeauthoryear{Endres et al.}{2016}]{endres16}
Endres, C. P., Schlemmer, S., Schilke, P., Stutzki, J., \& M\"uller, H. S. P. 2016, Journal of Molecular Spectroscopy, 327, 95 

\bibitem[\protect\citeauthoryear{Fedoseev et al.}{2018}]{fedoseev18}
Fedoseev, G., Scir\'e, C., Baratta, G. A., \& Palumbo, M. E. 2018, MNRAS, 475, 1819F
 
\bibitem[Fredon \& Cuppen(2018)]{fredon18}
Fredon, A., \& Cuppen, H. M. 2018, PCCP, 20, 5569F

\bibitem[Friberg et al.(1988)]{friberg88}
Friberg, P., Hjalmarson, A., Madden, S. C., \& Irvine, W. M. 1988, A\&A, 195,
281

\bibitem[Galli et al.(2019)]{galli19}
Galli, P. A. B., et al. 2019, A\&A, 630A, 137G

\bibitem[Garrod et al.(2017)]{garrod17}
Garrod, R. T., Belloche, A., M\"uller, H. S. P., \& Menten, K. M. 2017, A\&A, 601A, 48G

\bibitem[Garrod et al.(2008)]{garrod08}
Garrod, R. T., Weaver, S. L. W., \& Herbst, E. 2008, \apj, 682, 283

\bibitem[Garrod et al.(2007)]{garrod07}
Garrod, R. T., Wakelam, V., \& Herbst, E. 2007, A\&A, 467, 1103G

\bibitem[\protect\citeauthoryear{Harju et al.}{2019}]{harju19}
Harju, J., et al.  2020, ApJ, 895, 101H

\bibitem[\protect\citeauthoryear{Hasegawa et al.}{1993}]{hasegawa93}
Hasegawa, T. I., \& Herbst, E. 1993, MNRAS, 263, 589H

\bibitem[Herbst \& van Dishoeck(2009)]{herbst09}
Herbst, E., \& van Dishoeck, E. F. 2009, ARA\&A, 47, 427H

\bibitem[Hickson et al.(2016)]{hickson16}
Hickson, K. M., Wakelam, V., \& Loison, J.-C. 2016, MolAs, 3, 1H

\bibitem[\protect\citeauthoryear{Holdship et al.}{2019}]{holdship19}
Holdship, J., et al. 2019, ApJ, 880, 138H

\bibitem[\protect\citeauthoryear{Hollis et al.}{2006}]{hollis06}
Hollis, J. M., Lovas, F. J., Remijan, A. J., Jewell, P. R., Ilyushin, V. V., \& Kleiner, I. 2006, ApJ, 643L, 25H

\bibitem[\protect\citeauthoryear{Hollis et al.}{2000}]{hollis00}
Hollis, J. M., Lovas, F. J., \& Jewell, P. R. 2000, ApJ, 540L, 107H

\bibitem[Ivlev et al.(2015)]{ivlev15}
Ivlev, A. V., R\"ocker, T. B., Vasyunin, A., \& Caselli, P. 2015, ApJ, ApJ, 805, 59

\bibitem[\protect\citeauthoryear{Jim\'enez-Serra et al.}{2016}]{jimenez16}
Jim\'enez-Serra, I., et al., 2016, ApJ, 830L, 6J 

\bibitem[\protect\citeauthoryear{Jin \& Garrod}{2020}]{jin20}
Jin, M., \& Garrod, R. T. 2020, ApJS, accepted. arXiv:2006.11127

\bibitem[\protect\citeauthoryear{Jorgensen et al.}{2012}]{jorgensen12}
Jorgensen, J. K., Favre, C., Bisschop, S. E., Bourke, T. L., van Dishoeck, E. F., \& Schmalzl, M. 2012, ApJ, 757L, 4J

\bibitem[Lattanzi et al.(2020)]{lattanzi20}
Lattanzi, V., et al.  2020, A\&A, 633A, 118L

\bibitem[Marcelino et al.(2007)]{marcelino07}
Marcelino, N., Cernicharo, J., Ag\'undez, M., et al. 2007, ApJ, 665, L127

\bibitem[\protect\citeauthoryear{Mart\'{\i}n et al.}{2019}]{martin19}
Mart\'{\i}n, S., Mart\'{\i}n-Pintado, J., Blanco-S\'anchez, C., Rivilla, V. M., Rodr\'{\i}guez-Franco, A., \& Rico-Villas, F. 2019, A\&A, 631A, 159M

\bibitem[\protect\citeauthoryear{Minissale et al.}{2016}]{minissale16}
Minissale, M., Moudens, A., Baouche, S., Chaabouni, H., \& Dulieu, F. 2016, MNRAS, 458, 2953M

\bibitem[\protect\citeauthoryear{Nguyen et al.}{2018}]{nguyen18}
T. Nguyen1, S. Baouche1, E. Congiu1, S. Diana1, L. Pagani2 and F. Dulieu1

\bibitem[\"Oberg et al.(2010)]{oberg10}
\"Oberg, K. I., Bottinelli, S., Jorgensen, J. K., \& van Dishoeck, E. F. 2010, ApJ, 716, 825O

\bibitem[\protect\citeauthoryear{Pickett et al.}{1998}]{pickett98}
Pickett, H. M., Poynter, R. L., Cohen, E. A., et al. 1998, JQSRT, 60, 883 

\bibitem[\protect\citeauthoryear{Punanova et al.}{2018}]{punanova18}
Punanova, A., et al. 2018,  ApJ, 855, 112P

\bibitem[Quenard et al.(2018)]{quenard18}
Qu\'enard, D., Jim\'enez-Serra, I., Viti, S., Holdship, J., \& Coutens, A. 2018, MNRAS, 474, 2796Q

\bibitem[Redaelli et al.(2019)]{redaelli19}
Redaelli, E., Bizzocchi, L., Caselli, P., Sipil\"a, O., Lattanzi, V., Giuliano, B. M., \& Spezzano, S.  2019, A\&A, 629A, 15R

\bibitem[Rawlings et al.(2013)]{rawlings13}
Rawlings, J. M. C., Williams, D. A., Viti, S., \& Cecchi-Pestellini, C. 2013, MNRAS, 430, 264

\bibitem[Ruaud et al.(2015)]{ruaud15}
Ruaud, M., Loison, J. C., Hickson, K. M., Gratier, P., Hersant, F., \& Wakelam, V. 2015, MNRAS, 447, 4004R

\bibitem[\protect\citeauthoryear{Scibelli \& Shirley}{2020}]{scibelli20}
Scibelli, S., \& Shirley, Y. 2020, ApJ, 891, 73S

\bibitem[\protect\citeauthoryear{Shingledecker et al.}{2018}]{shingledecker18}
Shingledecker, C. N., Tennis, J., Le Gal, R., \& Herbst, E. 2018, ApJ, 861, 20S

\bibitem[\protect\citeauthoryear{Soma et al.}{2018}]{soma18}
Soma, T., Sakai, N., Watanabe, Y., \& Yamamoto, S. 2018, ApJ, 854, 116S

\bibitem[\protect\citeauthoryear{Spezzano et al.}{2017}]{spezzano17}
Spezzano, S., Caselli, P., Bizzocchi, L., Giuliano, B. M., \& Lattanzi, V. 2017, A\&A, 606A, 82S

\bibitem[\protect\citeauthoryear{Spezzano et al.}{2016}]{spezzano16}
Spezzano, S., Bizzocchi, L., Caselli, P., Harju, J., \& Br\"unken, S. 2016, A\&A, 592, L11

\bibitem[\protect\citeauthoryear{Tafalla et al.}{2002}]{tafalla02}
Tafalla, M., Myers, P. C., Caselli, P., Walmsley, C. M., \& Comito, C. 2002, ApJ, 569, 815T

\bibitem[\protect\citeauthoryear{Tafalla et al.}{2004}]{tafalla04}
Tafalla, M., Myers, P. C., Caselli, P., \& Walmsley, C. M.  2004, Ap\&SS, 292, 347T

\bibitem[\protect\citeauthoryear{Tafalla et al.}{2006}]{tafalla06}
Tafalla, M., Santiago-Garc\'{\i}a, J., Myers, P. C., Caselli, P., Walmsley, C. M., \& Crapsi, A. 2006, A\&A, 455, 577T

\bibitem[\protect\citeauthoryear{van der Tak et al.}{2007}]{vandertak07}
Van der Tak, F.F.S., Black, J.H., Sch\"oier, F.L., Jansen, D.J., van Dishoeck, E.F. 2007, A\&A, 468, 627

\bibitem[\protect\citeauthoryear{Vastel et al.}{2019}]{vastel19}
Vastel, C., Loison, J. C., Wakelam, V., \& Lefloch, B. 2019, A\&A, 625A, 91V

\bibitem[\protect\citeauthoryear{Vastel et al.}{2014}]{vastel14}
Vastel, C., Ceccarelli, C., Lefloch, B., \& Bachiller, R. 2014, ApJ, 795L, 2V

\bibitem[Vasyunin et al.(2017)]{vasyunin17}
Vasyunin, A. I., Caselli, P., Dulieu, F., \& Jim\'enez-Serra, I. 2017, ApJ, 842, 33V

\bibitem[Vasyunin \& Herbst(2013)]{vasyunin13}
Vasyunin, A. I., \& Herbst, E. 2013, ApJ, 769, 34	

\bibitem[Wakelam \& Herbst(2008)]{wakelam08}
Wakelam, V., \& Herbst, E. 2008, ApJ, 680, 371W

\bibitem[Watanabe \& Kouchi(2002)]{watanabe02}
 Watanabe, N., \& Kouchi, A. 2002, ApJ, 571L, 173W 
 
\bibitem[Willis et al.(2020)]{willis20} 
Willis, E. R. et al. 2020, A\&A, 636A, 29W

\bibitem[Wirstr\"om et al.(2011)]{wirstrom11} 
Wirstr\"om, E. S., et al. 2011, A\&A, 533A, 24W

\bibitem[Yoshida et al.(2019)]{yoshida19} 
Yoshida, K., et al.  2019, PASJ, 71S, 18Y

\end{thebibliography}
\end{document}